\documentclass[aps, prl, reprint, superscriptaddress, amsmath, amssymb, floatfix, longbibliography]{revtex4-2}

\usepackage{bm}           % Bold math symbols
\usepackage{mathtools}    % Extended math features and tools
\usepackage{dsfont}       % Double-stroke fonts (for \mathds{1})
\usepackage{mathrsfs}     % Script fonts (\mathscr)
\usepackage{mathdots}     % Additional dots for matrices and arrays

\usepackage{braket}       % Dirac bra-ket notation

\usepackage{graphicx}
\usepackage{multirow}
\usepackage{array}
\usepackage{dcolumn}      % Align table columns on decimal point

\usepackage[dvipsnames]{xcolor} % xcolor with extended color names support
\usepackage[normalem]{ulem}     % Support for strikethrough (\sulem)
\usepackage{hyperref}           % Hyperlinks (should be loaded last)

\newcommand{\ketbra}[2]{\ket{#1}\!\bra{#2}}

\begin{document}

\title{Coherence Transfer in Quantum Networks}

\date{\today}

\author{Chun-Yang Lin}
\affiliation{Department of Engineering Science, National Cheng Kung University, Tainan 70101, Taiwan}
\affiliation{Center for Quantum Frontiers of Research $\&$ Technology, National Cheng Kung University, Tainan 70101, Taiwan}

\author{Yu-Cheng Li}
\affiliation{Department of Engineering Science, National Cheng Kung University, Tainan 70101, Taiwan}
\affiliation{Center for Quantum Frontiers of Research $\&$ Technology, National Cheng Kung University, Tainan 70101, Taiwan}

\author{Shih-Hsuan Chen}
\affiliation{Department of Engineering Science, National Cheng Kung University, Tainan 70101, Taiwan}
\affiliation{Center for Quantum Frontiers of Research $\&$ Technology, National Cheng Kung University, Tainan 70101, Taiwan}

\author{Sheng-Yan Sun}
\affiliation{Department of Engineering Science, National Cheng Kung University, Tainan 70101, Taiwan}
\affiliation{Center for Quantum Frontiers of Research $\&$ Technology, National Cheng Kung University, Tainan 70101, Taiwan}

\author{Ching-Jui Huang}
\affiliation{Department of Engineering Science, National Cheng Kung University, Tainan 70101, Taiwan}
\affiliation{Center for Quantum Frontiers of Research $\&$ Technology, National Cheng Kung University, Tainan 70101, Taiwan}

\author{Kuan-Jou Wang}
\affiliation{Department of Engineering Science, National Cheng Kung University, Tainan 70101, Taiwan}
\affiliation{Center for Quantum Frontiers of Research $\&$ Technology, National Cheng Kung University, Tainan 70101, Taiwan}

\author{Che-Ming Li}
\email{cmli@mail.ncku.edu.tw}
\affiliation{Department of Engineering Science, National Cheng Kung University, Tainan 70101, Taiwan}
\affiliation{Center for Quantum Frontiers of Research $\&$ Technology, National Cheng Kung University, Tainan 70101, Taiwan}

\begin{abstract}

Detecting coherence transfer in complex quantum networks is challenging under uncharacterized experimental conditions and with limited access to the system. Here, we use the coexistence of static and dynamic coherence to derive a nonlinear criterion for coherence transfer. The criterion applies even when the verification capabilities of checkpoint nodes are uncharacterized. It requires only minimal one-network output analysis, up to the number of output state dimensions. Each output requires only two-setting measurements of state populations, regardless of the network size. The criterion is general for both spatial and temporal networks, including those with multiple coherence sources, and applies to quantum transport in solid-state nanostructures. Experimentally, using entanglement swapping, we transfer polarization-entangled pairs in four- and six-photon entanglement networks. The criterion provides experimental evidence of coherence transfer in multiphoton entanglement networks. Our findings provide a practical tool for investigating the presence, role, or function of coherence transfer in quantum networks.

\end{abstract}

\maketitle

\textit{Introduction.} Quantum networks~\cite{Kimble08,Wehner18}  are essential for exploring how quantum states transfer between diverse systems to enable nonclassical functions, from naturally occurring \textit{in vivo} energy transport~\cite{Lambert13}, to the engineering of a world clock~\cite{Komar14}, and from chip-scale requirements in quantum computers~\cite{Ladd10} to applications across the large-scale quantum internet~\cite{Gisin07,Ritter12}. A key feature is their ability to transfer quantum coherence between network nodes, a phenomenon classical physics cannot explain~\cite{Streltsov17}. This process relies on establishing and maintaining quantum coherence~\cite{Gour19,Liu20,Kuo19} and becomes a vital aspect of quantum networks as quantum techniques advance~\cite{Sne25}.

 Verifying coherence transfer in action, whether in biological~\cite{Jha26} or artificial~\cite{Lambert10,Biggerstaff16} systems, is challenging due to two experimental constraints: (1) uncharacterized experimental conditions and (2) restricted system access. Experimental trust issues stem from environmental noise and experimental flaws, causing a node's actual state to deviate from expectations, as in long-distance communication networks~\cite{Lu22,Yang25}. A node and its associated manipulations are considered uncharacterized when insufficient information exists about the noise and flaws. Access issues involve measuring or controlling the network and its components. Because a node's degrees of freedom may permit only basic or indirect measurements given the limited available techniques, this limited access leads to an incomplete understanding of networks, for example, quantum process tomography of charge transport through nanostructures using only available charge detection~\cite{Brandes05} and of the transfer of electronic excitations in a photosynthetic complex~\cite{Joel11} using two-dimensional electronic spectroscopy (2DES)~\cite{Collini21}.
 
 %The former struggles because only charge detection is available, whereas the latter faces challenges in two-dimensional electronic spectroscopy (2DES)~\cite{Collini21}, including state preparation and isotropic orientational averaging.

Progress has been made in understanding coherence~\cite{Baumgratz14,Streltsov17}, but confirming its transfer under both constraints remains difficult. While theories involving quantum operations, such as quantum channel resources~\cite{Gour19,Liu20} and quantum process capability~\cite{Kuo19}, analyze coherence, they require known quantum state and process tomography, which conflicts with constraints (1) and (2). Quantum coherence witnesses~\cite{Li12,Kofler13} can verify the quantum nature under constraint (2) but are incompatible with constraint (1) because they require measurements of propagators conditioned on precise state preparation and time control. Although the robustness of coherence witnesses~\cite{Napoli16,Piani16,Zheng18,Ma21,Li24} simplifies experiments, measurement errors can still affect their accuracy. They may help under constraint (2) but are vulnerable under constraint (1), depending on measurement precision. Testing the Leggett-Garg inequality~\cite{Leggett85,Knee12} can reveal quantum dynamics such as coherence creation or preservation under constraint (1). It requires non-invasive measurements, making practical application under constraint (2) difficult. Measurement-device-independent qubit coherence methods~\cite{Nie19} using tomography have been proposed for untrusted devices, potentially addressing constraint (1). However, they are limited by constraint (2) due to their reliance on tomography. Recently, quantum coherence in triangular networks with independent sources and joint measurements was demonstrated without adjustments~\cite{Bibak24}. While partially addressing constraints (1) and (2), their scope is limited to these specific structures and does not cover broader constraints.

\begin{figure}[t]
	\includegraphics[width=8.7cm]{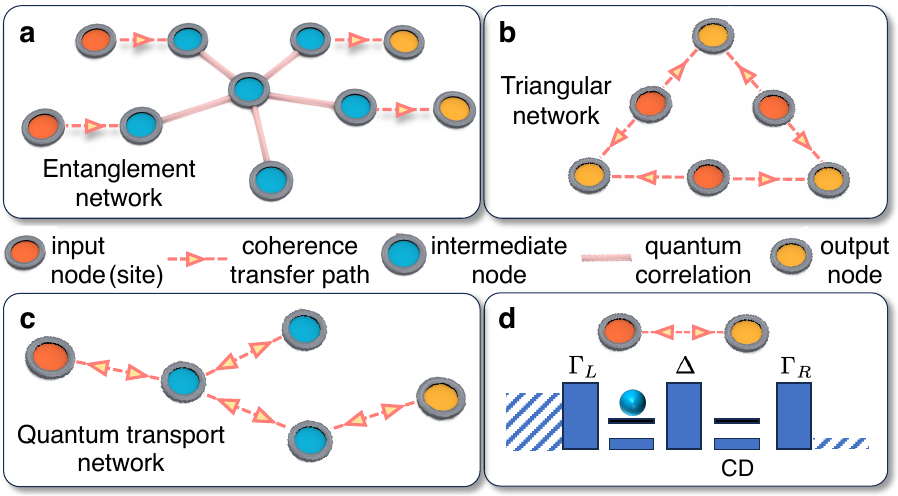}
	\caption{Coherence transfer in quantum networks. The coherence-transfer criterion (\ref{criterion}) and its temporal analog~(\ref{criterion2}) apply across various access levels and scenarios, including (a) entanglement networks~\cite{Pan12,Pompili21}, as demonstrated in our experiments; (b) the triangular network with three coherence sources~\cite{Bibak24}, where each input node emits a single particle in a superposition of two paths toward the output nodes; (c) quantum transport networks~\cite{Brandes05}, such as single-charge transport between sites (nodes) in nanostructures; and (d) a double quantum dot~\cite{Lambert10} serving as a primitive unit, with charge detectors (CDs) used for measurement.}
	\label{network}
\end{figure}

This work has a twofold goal. First, we introduce a practical and general coherence-transfer criterion that accommodates the experimental constraints (1) and (2). The criterion works under constraint (1) for nodes with unknown detection capabilities and, under constraint (2), requires only two measurement settings, irrespective of network size, for a minimal one-network output analysis, up to the number of output state dimensions. These two features make the criterion broadly applicable across various spatial and temporal coherence-transfer scenarios, including entanglement, multicoherence sources, and quantum transport networks. Second, we apply this criterion to the transfer of qubits within four- and six-photon entanglement networks and experimentally detect their coherence, which cannot be achieved with known methods under the limited experimental constraints~\cite{Gour19,Liu20,Kuo19,Li12,Kofler13,Napoli16,Piani16,Zheng18,Ma21,Li24,Leggett85,Knee12,Nie19,Bibak24}. These results help ease the difficulties encountered in coherence-transfer experiments~\cite{Lambert13,Komar14,Ladd10,Gisin07,Ritter12,Jha26,Lambert10,Biggerstaff16,Lu22,Yang25,Brandes05,Joel11,Collini21}.

\textit{Theory and methods.} Coherence transfer in networks, $\mathcal{E}_{\text{Net}}$, involves three stages: input, evolution, and output, with nodes either entangled~\cite{Pan12,Pompili21} or not~\cite{Awschalom21}. The input process sets the coherence of the transfer state, $\rho_{\text{in},k}$, at the input nodes, where $k$ labels the different states. During evolution, coherence propagates from the input through intermediate nodes to the output nodes. The output nodes then receive and output the state, $\rho_{\text{out},k}$, with coherence for applications such as measurement-based quantum information in entanglement networks~\cite{Raussendorf03,Barz12,Knaut24}. Figures~\ref{network}(a)-\ref{network}(c) illustrate the three stages across three scenarios.

\begin{figure}[t]
	\includegraphics[width=8.7cm]{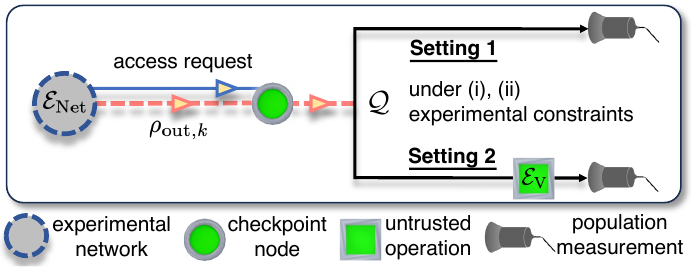}
	\caption{Coherence-transfer detection access system. Once the checkpoint node approves the coherence-detection task from the network ($\mathcal{E}_{\text{Net}}$), coherence-transfer detection determines $\mathcal{Q}$ [Eq.~(\ref{criterion})] by first measuring the output-state population $\rho_{\text{out},k}$ at the output checkpoint node (Setting~1) and then performing additional measurement rounds following an uncharacterized (untrusted) checkpoint operation $\mathcal{E}_\text{V}$ (Setting~2).}
	\label{scheme}
\end{figure}

We introduce a coherence-transfer detection access system, as shown in Fig.~\ref{scheme}. The experimental network's output node (not shown) and a checkpoint node with a verifier serve as the user and the access control manager, respectively. The checkpoint node first assesses whether characterizing $\mathcal{E}_{\text{V}}$ is feasible under the experimental constraints (1) and (2) and whether the output node can comply with the access policies and bear the cost of characterizing $\mathcal{E}_{\text{V}}$~\cite{cost}. If characterizing $\mathcal{E}_{\text{V}}$ is feasible and the output node can provide the required cost, the checkpoint node uses standard coherence measures (not shown in Fig.~\ref{scheme}). Otherwise, the checkpoint node implements coherence detection using our introduced criterion~\cite{cost}.

The verifier at the checkpoint nodes can follow the procedure below to detect coherence transfer. First, the verifier measures $\rho_{\text{out},k}$ to determine their \emph{static} populations in the experimental state basis, $\{\ket{i}\}=\{\ket{i}|i=0,1,...,d-1\}$, which form the diagonal matrix of the output states' density operator, $\rho_{\text{out},k}^{(\text{d})}$ (Setting~1 in Fig.~\ref{scheme}). Then, the verifier applies an operation, $\mathcal{E}_\text{V}$, with an unknown \emph{dynamic} ability to create coherence, on $\rho_{\text{out},k}$. Next, the verifier measures them to determine their populations, assuming a faithful measurement (Setting~2 in Fig.~\ref{scheme}). These populations are the probabilities of observing the basis states, $\ket{i}$ for the state $\mathcal{E}_\text{V}(\rho_{\text{out},k})$, denoted as $P_{\mathcal{E}_\text{V}}(i|\rho_{\text{out},k})$. Based on the experimental outcomes above, we identify a new property of coherence transfer that emerges only when both the output states of a network possess coherence and the operation applied to them can create coherence (dynamic coherence). With the coexistence of two coherences, we define our criterion for coherence transfer for the output states produced from the $n$ input-state types as a nonlinear form:
\begin{equation}
\mathcal{Q}\!:=\!\min_{\mathcal{E}_{\mathcal{I}}}\!\sum_{k=1}^{n}\sum_{i=0}^{d-1}\!\left|P_{\mathcal{E}_\text{V}}(i|\rho_{\text{out},k})\!-\!P_{\mathcal{E}_{\mathcal{I}}}(i|\rho_{\text{out},k}^{(\text{d})})\right|\!>\!0,\label{criterion}
\end{equation}
where $P_{\mathcal{E}_{\mathcal{I}}}(i|\rho^{(\text{d})}_{\text{out},k})=\text{tr}[\ket{i}\!\!\bra{i}\mathcal{E}_{\mathcal{I}}(\rho^{(\text{d})}_{\text{out},k})]$ and $\mathcal{E}_{\mathcal{I}}$ is a process that cannot generate coherence from any incoherent states~\cite{Kuo19,SM}. The kernel $\mathcal{Q}$ is obtained by minimizing over all incapable processes and can be computed using semidefinite programming (SDP)~\cite{Lofberg2004Yalmip,Toh1999Sdpt3,Andersen2000Mosek,SM}.

When criterion (\ref{criterion}) is satisfied, the process $\mathcal{E}_{\text{Net}}$ is confirmed to transfer coherence effectively, and $\mathcal{E}_\text{V}$ is recognized as capable of generating coherence. Its detection resolution depends primarily on the assumption of reliable population measurements and on the coherence properties of $\rho_{\text{out},k}$ and $\mathcal{E}_{\text{V}}$. The tightness, optimality, detection power, error in a measured $\mathcal{Q}$, and SDP numerical errors related to the kernel or criterion are detailed in the Supplemental Material~\cite{SM}. Because criterion~(\ref{criterion}) captures both the static coherence feature and the dynamic transfer property, it is comprehensive and suitable for a wide range of coherence network detection tasks under experimental constraints~\cite{compare}, compared with existing work~\cite{Gour19,Liu20,Kuo19,Li12,Kofler13,Napoli16,Piani16,Zheng18,Ma21,Li24,Leggett85,Knee12,Nie19,Bibak24}. It is worth noting that when it is experimentally feasible to characterize and further calibrate $\mathcal{E}_{\text{V}}$, it can be chosen to maximize the criterion based on $\mathcal{E}_{\text{V}}$'s coherence-creation capability, thereby yielding the maximum value of $\mathcal{Q}$~\cite{SM}.

To prove criterion~(\ref{criterion}), we use the operator-sum representation~\cite{Chuang97} to model $\mathcal{E}_{\text{V}}$ acting on $\rho_{\text{out},k}=\sum_{m,n=0}^{d-1}\rho_{k,mn}E_{mn}$, with $E_{mn}=\ketbra{m}{n}$, as $\mathcal{E}_{\text{V}}(\rho_{\text{out},k})=\sum_{m,n,i,l=0}^{d-1}\chi_{mn,il}E_{mn}\rho_{\text{out},k}E_{il}^\dag$, where $\chi_{mn,il}$ is the process matrix of $\mathcal{E}_{\text{V}}$ and describes the transition from the coherence terms $\ketbra{n}{l}$ to the diagonal elements $\ketbra{m}{m}$ when $m=i$ and $n\neq l$. Then, we have $P_{\mathcal{E}_\text{V}}(i|\rho_{\text{out},k})=\sum_{\substack{n,l=0, n \ne l}}^{d-1} \rho_{k,nl} \chi_{in,il}+\sum_{n=0}^{d-1} \rho_{k,nn} \chi_{in,in}$ and $P_{\mathcal{E}_\mathcal{I}}(i|\rho^{(\text{d})}_{\text{out},k})=\sum_{n=0}^{d-1} \rho_{k,nn} \chi_{\mathcal{I},in,in}$, where $\chi_{\mathcal{I},in,in}$ is the process matrix element of $\mathcal{E}_\mathcal{I}$~\cite{Kuo19}. These imply that $\mathcal{Q} = 0$ when $\rho_{k,nl}=0$ and $n\neq l$ (the network process cannot generate coherence), or when $\chi_{in,il} = 0$ and $n \neq l$ (the verifier cannot produce coherence), or both conditions are met simultaneously. Therefore, $\mathcal{Q} > 0$ clearly indicates that the network process can transfer coherence and that $\mathcal{E}_{\text{V}}$ can generate coherence, as described by Eq.~(\ref{criterion})~\cite{iff}.

The two-setting-only criterion generally applies to scenarios where population measurements and an uncharacterized $\mathcal{E}_\text{V}$ are available for analyzing sufficient network output states. It is summarized in the three groups below.

(G1) $n=1$. When a network is defined within a specific, well-understood physical framework, $\mathcal{E}_{\mathcal{I}}$ can be directly described by that framework. That is, we require precise \textit{a priori} knowledge of the network structure and the quantum-mechanical descriptions, as well as the physical assumptions underlying the absence of network coherence-transfer capability and the checkpoint's lack of coherence-creation capability. Together, these two requirements make $\mathcal{E}_{\mathcal{I}}$ explicitly suited to a concrete network, so minimizing the kernel $\mathcal{Q}$ is analytically feasible for defining the criterion's operational boundaries~(\ref{criterion}). The triangular network with three coherence sources~\cite{Bibak24} [Fig.~\ref{network}(b)] is a concrete example that illustrates how these requirements make the $n=1$ case applicable. To detect the coherence transfer of a joint state, $\rho_{\text{out}}$, in the network generated when each input node emits a single particle in a superposition of two paths toward the output nodes, one can use a 50:50 beam splitter (BS) at each vertex of the triangle to interfere the two incoming modes, acting as the checkpoint operation $\mathcal{E}_{\text{V}}$, and consider the following criterion:
\begin{equation}
	\mathcal{Q} \!=\! \min_{\mathcal{E}_{\mathcal{I}}}\!\!\!\!\sum_{a,b,c=0}^{1}
	\!\!\left | P_{\mathcal{E}_{\text{V}}}(a,\!b,\!c \mid \!\rho_{\text{out}})
	\!-\!P_{\mathcal{E}_{\mathcal{I}}}(a,\!b,\!c\mid \!\rho^{(\text{d})}_{\text{out}}) \right|\!>\!0,\label{tri}
\end{equation}
where $a$, $b$, and $c$ represent the binary postselected measurement results for each BS's two output modes. The kernel is analytically minimized by considering all incapable processes within a classical, realistic model that cannot yield coherence, thereby eliminating the need for an additional experiment to obtain $\rho^{(\text{d})}_{\text{out}}$. Therefore, $\mathcal{Q} > 0$ holds in this scenario, implying that coherence transfer can be detected without changing the state or modifying the local measurement setting~\cite{Bibak24}. See the Supplemental Material for further discussion~\cite{SM}. Compared with the existing detection method~\cite{Bibak24}, criterion~(\ref{tri}) identifies the role of BSs and shows that their operation can produce coherence. The conclusion above applies directly to networks with multiple vertices and coherence sources.

(G2) $n\geq2$. In a broader context, when knowledge of the network is limited, $\mathcal{E}_{\mathcal{I}}$ is defined by its failure to achieve coherence within the general quantum operations formalism~\cite{Kuo19}. To be practical in these cases, as demonstrated experimentally below for entanglement networks, at least two distinct input and output state configurations are required ($n \ge 2$). See the Supplemental Material~\cite{SM} for a detailed discussion.

\begin{figure}[t]
	\includegraphics[width=8.7cm]{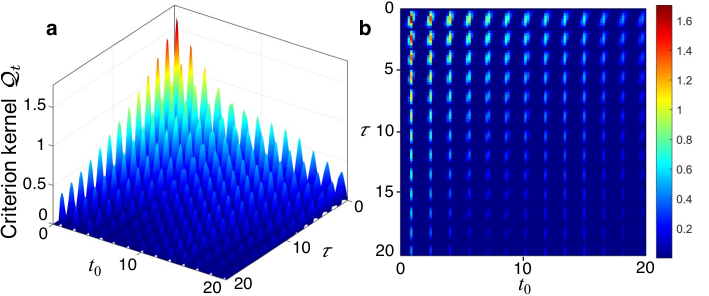}
	\caption{Detecting coherence transfer in single-electron transport through a double quantum dot. For the setup in Fig.~\ref{network}(d), we assume weak coupling, large bias, and Coulomb blockade, with CDs used for the measurements. The left and right tunneling rates and the coherent tunneling amplitude between the left and right dots are $\Gamma_{L}=4$, $\Gamma_{R}=0.1$, and $\Delta=1$, respectively. (a) The nonvanishing $\mathcal{Q}_{t}$ indicates regions where coherence develops during quantum transport, as shown in its density plot [pannel (b)], where $\tau=t-t_{0}$ is the time difference for Eq.~(\ref{criterion2}). See the Supplemental Material~\cite{SM} for a detailed discussion, including the SDP complexity.}
	\label{DQD}
\end{figure}

\begin{figure*}[t]
\includegraphics[width=17cm]{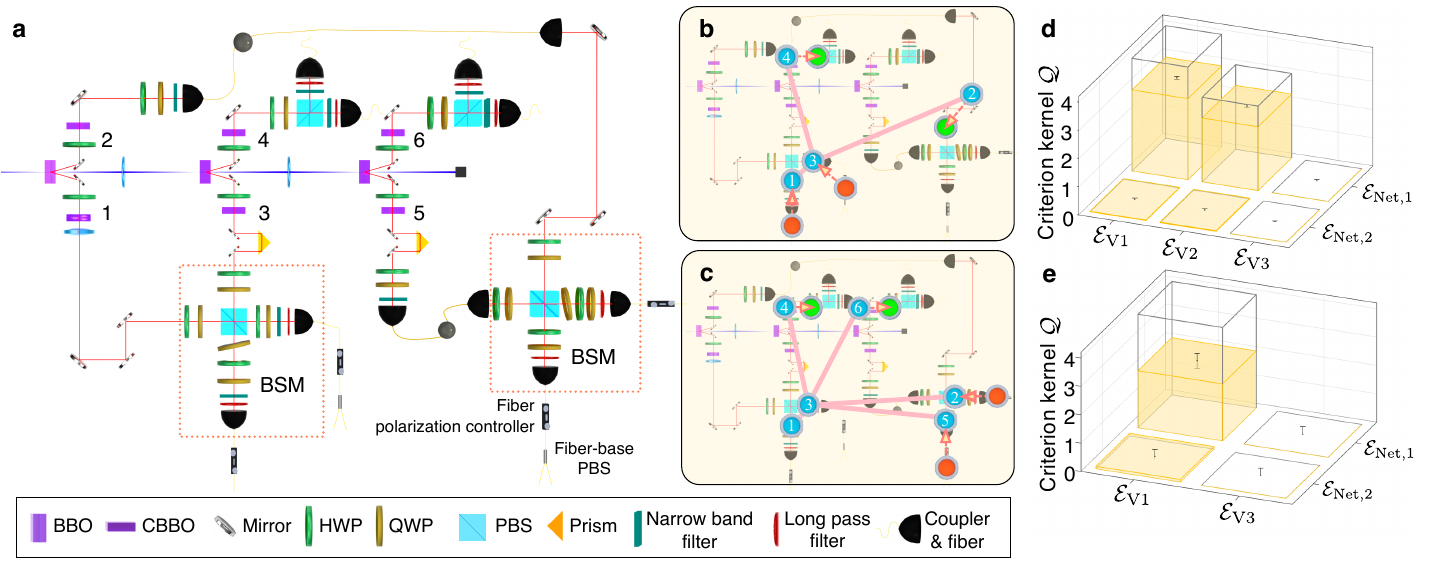}
\caption{Detection of experimental coherence transfer in multi-photon entanglement networks. (a) Experimental setup. The three spontaneous parametric down-conversion (SPDC) sources~\cite{Pan12} are pumped by a pulsed laser beam. In each source, a $\beta$-barium-borate (BBO) crystal is pumped to generate pairs of polarization-entangled photons via type-II non-collinear SPDC, with walk-off compensation provided by an additional BBO crystal (CBBO). The three photon pairs prepared in modes 1-2, 3-4, and 5-6 are intended to generate the target state $\left|\psi^-\right\rangle=(1/\sqrt{2})(\left|HV\right\rangle-\left|VH\right\rangle)$, where $\left|H\right\rangle$ ($\left|V\right\rangle$) denotes the horizontal (vertical) photon polarization state. Combined with wave-plate sets, two photonic fusion units temporally and spatially overlap photon pairs at a PBS to perform either a photonic fusion operation~\cite{Pan98,Lu07} or a BSM~\cite{Bouwmeester97,Pan98}. The results can be considered as star multiphoton entanglement networks [Fig.~\ref{network}(a)]. Two-qubit coherence transfer pathways in the four-photon and six-photon networks are shown in panels (b) and (c), respectively, with corresponding experimental detection results in panels (d) and (e), identified by criterion~(\ref{criterion}). See the Supplemental Material \cite{SM} for the experimental details and results above.}
	\label{fig:expt}
\end{figure*}

(G3) Networks in the time domain. For an unknown dynamics $\mathcal{E}_{t,0}$ in an open quantum system coupled to a reservoir~\cite{Breuer16}, criterion~(\ref{criterion}) has a temporal version that can be used to determine whether it can generate coherence in the state basis $\{\ket{i}\}$ over time, for example, in the site basis of quantum transport networks~[Fig.~\ref{network}(c)]. Initially, we consider $n$ states with different initial conditions at time 0, denoted by $\rho_{k,0}$, for $k=1,2,..., n$. For each $k$, we perform population measurements at time $t_0$ on the states $\rho_{k,t_0}$ that have evolved under the dynamics $\mathcal{E}_{t_0,0}$ from time 0 to $t_0$. Subsequently, we also conduct population measurements at a later time $t$, where $t > t_0$, on the states $\rho_{k,t}$ that have evolved under the dynamics $\mathcal{E}_{t,0}$ from 0 to $t$. This includes the additional evolution from $t_0$ to $t$, which is described by the dynamics $\mathcal{E}_{t,t_0}$, corresponding to the role of $\mathcal{E}_{\text{V}}$ in Eq.~({\ref{criterion}). We then determine the probability of observing the state $\ket{i}$ at time $t$: $P(i|\rho_{k,t})$, and the diagonal component of $\rho_{k,t_0}$: $\rho_{k,t_0}^{(\text{d})}$. The temporal equivalent of criterion~(\ref{criterion}) states that:
\begin{equation}
\mathcal{Q}_{t}:=\min_{\mathcal{E}_{\mathcal{I}}}\sum_{k=1}^{n}\!\sum_{i=0}^{d-1}\left|P(i|\rho_{k,t})-P_{\mathcal{E}_{\mathcal{I}}}(i|\rho_{k,t_0}^{(\text{d})})\right|>0,\label{criterion2}
\end{equation}
where $P_{\mathcal{E}_{\mathcal{I}}}(i|\rho^{(\text{d})}_{k,t_0})=\text{tr}[\ket{i}\!\!\bra{i}\mathcal{E}_{\mathcal{I}}(\rho^{(\text{d})}_{k,t_0})]$. For systems coupled to the reservoir Markovianly~\cite{Breuer16}, any dynamics that cannot produce coherent transitions between states in the basis $\{\ket{i}\}$ yield $\mathcal{Q}_{t}=0$~\cite{SM}. When experimental results satisfy criterion~(\ref{criterion2}), $\mathcal{E}_{t,0}$ is confirmed to generate coherence between states over time.

Criterion~(\ref{criterion2}) shares the same advantages as criterion~(\ref{criterion}) in the time domain. First, compared with the most effective time-domain witness~\cite{Li12,Kofler13}, our criterion is more practical because it does not require additional reliable reference states or propagator measurements. Second, the role of $\mathcal{E}_{\text{V}}$ is played by the additional system dynamics from $t_0$ to $t$. This makes detection of coherence dynamics experimentally feasible using only population measurements. Using quantum transport in the Markovian limit as an example [Figs.~\ref{network}(c) and~\ref{network}(d)], coherence dynamics can be detected solely from site population measurements of single charges [Fig.~\ref{network}(d)]. Finally, criterion~(\ref{criterion2}) is highly sensitive. As shown in Fig.~\ref{DQD}, it detects coherence in single-electron transport through a double quantum dot. Compared with the time periods identified by violations of the one-initial-state, three-time-point Leggett-Garg-type inequality~\cite{Lambert10} (which a classical Markovian system under noninvasive charge measurement satisfies), our criterion can detect a much broader window of quantum coherence, given three initial states ($n=3$) for considering $\mathcal{Q}$ and implementing minimization via SDP, and within the same limits to which the master equation formalism applies~\cite{Lambert10}. This identifies coherence transfer through the coexistence of static and dynamic coherence, distinct from approaches that rule out a classical description of transport through a nanostructure using a Leggett-Garg-type inequality.

\textit{Experiments.} Building on our previous results~\cite{Kao24}, we demonstrate that criterion~(\ref{criterion}) detects one- and two-qubit coherence transfer in four- and six-photon polarization-entanglement networks across four transmission scenarios. Two-qubit transfer involves inputting a Bell state and transferring it via a Bell-state measurement (BSM)~\cite{Bouwmeester97,Pan98} using wave plates, a polarizing beam splitter (PBS), and shared entanglement [see Fig.~\ref{fig:expt}(a)]. In four-photon networks [Fig.~\ref{fig:expt}(b)], coherence propagates from node~1 and node~3 to node~2 and node~4 through a single BSM. In six-photon networks [Fig.~\ref{fig:expt}(c)], the coherence input occurs via a BSM from node~2 and node~5, assisted by the BSM of node~1 and node~3, to node~4 and node~6. This process can also be viewed as entanglement swapping~\cite{Zukowski93,Pan98ES} within a star six-photon entanglement network [Fig.~\ref{network}(a)] using Bell states and BSMs. The effectiveness of coherence transfer, evaluated using $\mathcal{Q}$, is based on the basis $\{\ket{i}\}=\{\ket{0}=\ket{HH},\ket{1}=\ket{HV},\ket{2}=\ket{VH},\ket{3}=\ket{VV}\}$. In each multiphoton entanglement network, we perform two coherence-transfer processes, $\mathcal{E}_{\text{Net},1}$ and $\mathcal{E}_{\text{Net},2}$, designed to be \emph{capable} of achieving the target process above and \emph{incapable} of creating coherence under postselection~\cite{SM}, respectively. The operations applied to output states at the checkpoint node $\mathcal{E}_{\text{V}1}$, $\mathcal{E}_{\text{V}2}$, and $\mathcal{E}_{\text{V}3}$ correspond to quarter-wave plate (QWP) settings with the fast axis at $45^{\circ}$, $30^{\circ}$, and $0^{\circ}$ relative to the horizontal axis, respectively, while the half-wave plate (HWP) always sets the fast axis at $0^{\circ}$ relative to the horizontal axis. Evidence of coherence transfer and the effectiveness of $\mathcal{E}_{\text{V}}$ in four-photon and six-photon networks is shown in Figs.~\ref{fig:expt}(d) and~\ref{fig:expt}(e), respectively. Here, the measurement results are minimized relative to those from $\mathcal{E}_{\mathcal{I}}$ by using SDP~\cite{Lofberg2004Yalmip,Toh1999Sdpt3,Andersen2000Mosek,SM} to derive $\mathcal{Q}$. The ideal value of $\mathcal{Q}$ is shown by the gray solid line, while the experimentally measured value is depicted by the yellow bar. The difference between the ideal and experimental values primarily arises from limited state fidelity and photonic-fusion errors. See the Supplemental Material~\cite{SM} for details, including the relationship between criterion~(\ref{criterion}) and the witness for genuine multipartite entanglement.

Similarly, in faithful one-qubit coherence transfer, we first establish coherence and then transfer it via remote state preparation~\cite{Bennett01,Chen24}, wave plates, polarization measurements, and shared entanglement, followed by photonic fusion operations to subsequent stage nodes. See the Supplemental Material~\cite{SM} for details.

Our proof-of-principle experiment demonstrates the first identification of coherence transfer using two measurement settings. While all photonic elements required to determine $\mathcal{Q}$ can, in principle, be calibrated, the criterion in Eq.~(\ref{criterion}) allows the elements for $\mathcal{E}_{\text{V}}$ to remain uncharacterized under the prescribed assumption~\cite{assump}. This experimental demonstration helps clarify general quantum network scenarios in which uncharacterized situations can arise. In such cases, $\mathcal{E}_{\text{V}}$ is generally assumed to be uncharacterized or uncalibrated by default.

\textit{Conclusion and outlook.} In this work, we have developed a general criterion for detecting coherence transfers in quantum networks under typical experimental constraints. It requires only two measurement settings, regardless of the network’s size, even when checkpoint nodes are uncharacterized, and measures only a minimal one-network output, up to the number of output state dimensions. By identifying the coexistence of static and dynamic coherence, it is comprehensive and suitable for a wide range of coherence network detection tasks across different experimental access levels, from spatial to temporal, including entanglement, multicoherence sources, and quantum transport networks, compared with existing works. While existing methods have fallen short in assessing coherence under the experimental constraints, our experimentally derived values of the criterion kernel clearly demonstrate coherence transfer in the multiphoton entanglement networks we constructed. More generally, this fundamental yet practical framework helps explore coherence transfer in complex systems, such as transient coherences in photosynthetic systems, while accounting for practical constraints in 2DES.

\textit{Acknowledgments.} We thank N. Lambert, F. Nori, G.-Y. Chen, Y.-N. Chen, H. Lu, Y.-A. Chen, and J.-W. Pan for their valuable discussions on coherence, which motivated our research from previous work~\cite{Li12} to this current study. We also appreciate J.-Y. Wu and C.-S. Chuu for their helpful feedback. This work was partially supported by the National Science and Technology Council, Taiwan, under Grant Numbers NSTC 114-2112-M-006-016-MY3, NSTC 114-2119-M-007-012, and NSTC 115-2119-M-007-004.

\textit{Data Availability.} The theoretical and experimental data supporting the findings of this work are available upon reasonable request from the authors.

\bibliography{QCTrf}

\end{document}

\title{Supplementary Material for\\ Coherence Transfer in Quantum Networks}

\date{\today}

\author{Chun-Yang Lin}
\affiliation{Department of Engineering Science, National Cheng Kung University, Tainan 70101, Taiwan}
\affiliation{Center for Quantum Frontiers of Research $\&$ Technology, National Cheng Kung University, Tainan 70101, Taiwan}

\author{Yu-Cheng Li}
\affiliation{Department of Engineering Science, National Cheng Kung University, Tainan 70101, Taiwan}
\affiliation{Center for Quantum Frontiers of Research $\&$ Technology, National Cheng Kung University, Tainan 70101, Taiwan}

\author{Shih-Hsuan Chen}
\affiliation{Department of Engineering Science, National Cheng Kung University, Tainan 70101, Taiwan}
\affiliation{Center for Quantum Frontiers of Research $\&$ Technology, National Cheng Kung University, Tainan 70101, Taiwan}

\author{Sheng-Yan Sun}
\affiliation{Department of Engineering Science, National Cheng Kung University, Tainan 70101, Taiwan}
\affiliation{Center for Quantum Frontiers of Research $\&$ Technology, National Cheng Kung University, Tainan 70101, Taiwan}

\author{Ching-Jui Huang}
\affiliation{Department of Engineering Science, National Cheng Kung University, Tainan 70101, Taiwan}
\affiliation{Center for Quantum Frontiers of Research $\&$ Technology, National Cheng Kung University, Tainan 70101, Taiwan}

\author{Kuan-Jou Wang}
\affiliation{Department of Engineering Science, National Cheng Kung University, Tainan 70101, Taiwan}
\affiliation{Center for Quantum Frontiers of Research $\&$ Technology, National Cheng Kung University, Tainan 70101, Taiwan}

\author{Che-Ming Li}
\email{cmli@mail.ncku.edu.tw}
\affiliation{Department of Engineering Science, National Cheng Kung University, Tainan 70101, Taiwan}
\affiliation{Center for Quantum Frontiers of Research $\&$ Technology, National Cheng Kung University, Tainan 70101, Taiwan}

\begin{abstract}
In this supplementary information, we provide supporting material for the theoretical and experimental results discussed in the main text. First, in addition to the uncharacterized checkpoint node, we describe and discuss a scenario in which it can be characterized and optimized. We then explain the generality of the coherence-transfer criterion with respect to the required numbers of input and output state settings. Additionally, we provide the proof of the criterion used to detect coherence dynamics in open quantum systems and detail the time-domain example given in the main text for quantum transport in a double quantum dot. We also discuss the properties of the proposed criterion kernel $\mathcal{Q}$ and the criteria in Eqs.~(1) and~(3). Second, we present details of the coherence-transfer experiments in multi-photon entanglement networks. We start with entangled-photon sources and the generation of multi-photon entanglement. Then, we demonstrate how to use remote state preparation and entanglement swapping, including Bell state measurements, to implement the transfer of single-polarization qubits (not shown in the main text) and polarization-entangled pairs in four- and six-photon networks. We then present the details of measuring and analyzing the criterion kernels to provide experimental evidence of coherence-transfer processes in multi-photon entanglement networks. Finally, we discuss how the completed coherence-transfer experiments are generally representative of the key aspects of coherence transfer in entanglement networks. We also discuss the relationship between the coherence-transfer criterion and the entanglement witness.
\end{abstract}

\maketitle

\section{Characterized checkpoint operation $\mathcal{E}_{\text{V}}$}

Our framework allows the checkpoint ($\mathcal{E}_{\text{V}}$) to be either characterized or uncharacterized. In particular, this framework allows checkpoint nodes to be uncharacterized; that is, their capability to create coherence is unknown. We emphasize in the main text the importance of treating checkpoint nodes as uncharacterized in quantum networks. Although they are uncharacterized initially, once $\mathcal{Q}>0$ [Eq.~(1) in the main text], they can eventually be identified as capable of generating coherence in coherent network output states.

If $\mathcal{E}_{\text{V}}$ is assumed capable of generating coherence, the optimal $\mathcal{E}_{\text{V}}$ can be found as follows. To discuss how the checkpoint operation $\mathcal{E}_{\text{V}}$ should be chosen to optimize the criterion, we first revisit the scenario to which Eq.~(1) applies. The introduced framework allows checkpoint nodes to be either characterized or uncharacterized. When the assumption that knowledge of the checkpoint is required is relaxed, once $\mathcal{Q}>0$, an uncharacterized operation $\mathcal{E}_{\text{V}}$ can be identified as capable of creating coherence. On the other hand, if $\mathcal{E}_{\text{V}}$ is assumed capable of generating coherence, the optimal $\mathcal{E}_{\text{V}}$ can be obtained from the proof of Eq.~(1) in the main text. In that proof, we know that $P_{\mathcal{E}_\text{V}}(i|\rho_{\text{out},k})=\sum_{\substack{n,l=0, n \ne l}}^{d-1} \rho_{k,nl} \chi_{in,il}+\sum_{n=0}^{d-1} \rho_{k,nn} \chi_{in,in}$, and $P_{\mathcal{E}_\mathcal{I}}(i|\rho^{(\text{d})}_{\text{out},k})=\sum_{n=0}^{d-1} \rho_{k,nn} \chi_{\mathcal{I},in,in}$, where $\chi_{\mathcal{I},in,in}$ denotes the process matrix element of the process $\mathcal{E}_\mathcal{I}$ that is incapable of creating coherence. Therefore, the criterion kernel becomes:
\begin{eqnarray}
\mathcal{Q}&=&\min_{\mathcal{E}_{\mathcal{I}}}\!\sum_{k=1}^{n}\sum_{i=0}^{d-1}\!\left|P_{\mathcal{E}_\text{V}}(i|\rho_{\text{out},k})\!-\!P_{\mathcal{E}_{\mathcal{I}}}(i|\rho_{\text{out},k}^{(\text{d})})\right|\nonumber\\
&=&\!\sum_{k=1}^{n}\sum_{i=0}^{d-1}\!\left|\sum_{\substack{n,l=0, n \ne l}}^{d-1} \rho_{k,nl} \chi_{in,il}\right|\nonumber\\
&\leq&\!\sum_{k=1}^{n}\sqrt{\left(\sum_{\substack{n,l=0, n \ne l}}^{d-1} |\rho_{k,nl}|^2 \right)}\sum_{i=0}^{d-1}\sqrt{\sum_{\substack{n,l=0, n \ne l}}^{d-1}|\chi_{in,il}|^2},\nonumber\\ \label{r1}
\end{eqnarray}
where the second equality from the first is based on the setting $\chi_{\mathcal{I},in,in}=\chi_{in,in}$, and the third follows from applying the Cauchy–Schwarz inequality to the second.

If the optimality of $\mathcal{Q}$ is defined as the maximum value it can achieve when producing a coherent network output state, the above result shows that choosing a $\mathcal{E}_{\text{V}}$ that maximizes $\sum_{i=0}^{d-1}\sqrt{\sum_{\substack{n,l=0, n \ne l}}^{d-1}|\chi_{in,il}|^2}$ is necessary to achieve the optimality of $\mathcal{Q}$ (\ref{r1}), regardless of the coherent output state. Because the process matrix elements $\chi_{in,il}$ describe how a coherence-creation process induces transitions from coherence to population terms in a density matrix, the dynamic coherence features of $\chi$ revealed by the above derivation also clarify the important role of $\mathcal{E}_{\text{V}}$'s coherence-creation capability in detecting coherence transfer across networks.

While calibrating (characterizing) $\mathcal{E}_{\text{V}}$ helps achieve the optimal condition shown above, the calibration process can be challenging in general quantum networks due to two experimental constraints, namely (i) uncharacterized experimental conditions and (ii) restricted system access, as discussed in the introduction of the main text.

\section{Example of the single input and output state setting, $n=1$: triangular network} \label{app:triangular}

As more information about the network's internal structure becomes available, criterion~(1) in the main text naturally adapts into a more specific form, criterion~(2), that aligns with the given network architecture. A single input and output state setting, $n=1$, is sufficient for the criterion. As an explicit example of our proposed criterion within a specific physical framework, we consider the triangular network for quantum coherence introduced in Ref.~\cite{Bibak24}. See also Fig.~1(b) in the main text.

\begin{figure}[t]
	\includegraphics[width=8.8cm]{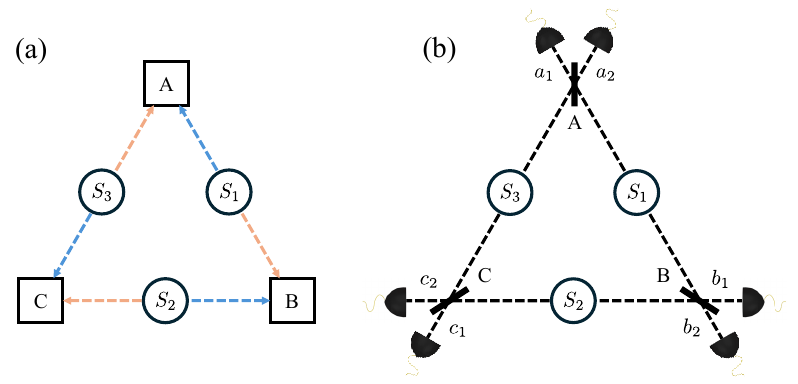}
	\caption{Three-party classical and quantum triangular networks. See also Fig.~1(b) in the main text. (a) In the triangle scenario with classical sources, particles are directed to the three measurement stations based on classical probabilities. In a postselected scenario, only two configurations remain: all sources send their particles clockwise (orange arrows) or counterclockwise (blue arrows). (b) In the triangular scenario with three quantum sources, each quantum source emits a single particle in a superposition of two propagation paths toward the stations (nodes) at the triangle's vertices. At each station, the two incoming modes interfere at a 50:50 beam splitter with two output detectors, which act as the operations of the checkpoint nodes, $\mathcal{E}_{\text{V}}$, in criterion~(1) in the main text. The analysis postselects the events in which exactly one particle is detected at each station.}
	\label{fig:triangular_network}
\end{figure}

According to Ref.~\cite{Bibak24}, three parties, Alice (A), Bob (B), and Claire (C), sit at the vertices of a triangle, and each pair of parties is connected by a single-particle source labeled $S_1$, $S_2$, and $S_3$, respectively, as shown in Fig.~\ref{fig:triangular_network}. In every run, each source emits one information carrier, and different sources act independently. Each party performs a fixed local measurement and outputs a binary result $a,b,c\in\{0,1\}$. The analysis is restricted to a postselected scenario in which only events with exactly one particle detected at each party are retained. The target is to realize a three-party distribution $p(a,b,c)$ satisfying the even-parity constraint:
\begin{equation}
	\begin{aligned}
		p(a,b,c \mid a \oplus b \oplus c = 0) &\neq 0
		\quad \text{and}\\
		p(a,b,c \mid a \oplus b \oplus c = 1) &= 0
		\quad \forall\, a,b,c.
	\end{aligned}
	\label{eq:target_constraint}
\end{equation}
The classical and quantum versions of the triangular network described in Ref.~\cite{Bibak24} are formulated within the following framework.

In the classical version, as shown in Fig.~\ref{fig:triangular_network}(a), each source routes its particle along a definite path with classical probabilities, so after postselection, only two global configurations remain: all particles going clockwise or all going counterclockwise. In this classical realistic model, the post-selected classical correlations take the form:
\begin{equation}
	p_{\mathrm{cl}}(a,b,c)= \gamma p_A(a) p_B(b) p_C(c)+ (1-\gamma) q_A(a) q_B(b) q_C(c),\label{eq:classical_model}
\end{equation}
where $0\le\gamma\le1$ is the probability of particles being distributed clockwise, and $p_X(x)$ [$q_X(x)$] is the probability that party $X\in\{A,B,C\}$ outputs $x\in\{0,1\}$ in the clockwise (counterclockwise) configuration. One finds that the even-parity constraint in Eq.~\eqref{eq:target_constraint} cannot be satisfied exactly by any choice of $\gamma$, $p_X$, and $q_X$.

In the quantum version, as shown in Fig.~\ref{fig:triangular_network}(b), each source emits a single particle coherently superposed over its two possible paths. Denoting by $S_{1A}^\dagger$ and $S_{1B}^\dagger$ the creation operators for the two output modes of source $S_1$, and analogously for $S_2$ and $S_3$, the initial state is:
\begin{equation}
	\ket{\psi_{\mathrm{in}}}
	=
	\left(\frac{1}{\sqrt{2}}\right)^{3}
	\bigl(S_{1A}^\dagger + S_{1B}^\dagger\bigr)
	\bigl(S_{2B}^\dagger + S_{2C}^\dagger\bigr)
	\bigl(S_{3C}^\dagger + S_{3A}^\dagger\bigr)\ket{0},
	\label{eq:path_superposition_initial_state}
\end{equation}
where $\ket{0}$ is the vacuum state. At each station (node), the two incoming modes interfere on a 50:50 beam splitter (BS), yielding output modes $(a_1^\dagger,a_2^\dagger)$ at Alice, $(b_1^\dagger,b_2^\dagger)$ at Bob, and $(c_1^\dagger,c_2^\dagger)$ at Claire, which acts as the checkpoint operation $\mathcal{E}_{\text{V}}$. After the BSs and postselection onto the subspace with exactly one particle per party, the normalized state becomes:
\begin{equation}
	\ket{\psi_{\mathrm{post}}}
	=
	\frac{1}{2}\Bigl[
	a_1^\dagger\bigl(b_1^\dagger c_1^\dagger + b_2^\dagger c_2^\dagger\bigr)
	+
	a_2^\dagger\bigl(b_2^\dagger c_1^\dagger + b_1^\dagger c_2^\dagger\bigr)
	\Bigr]\ket{0}.
	\label{eq:path_superposition_post_state}
\end{equation}
Each party records which output port was clicked, identifying detector 1 as having outcome $0$ and detector 2 as having outcome $1$. The resulting quantum probability distribution is:
\begin{equation}
	\begin{gathered}
		p_{\mathrm{Q}}(0,0,0)
		= p_{\mathrm{Q}}(0,1,1)
		= p_{\mathrm{Q}}(1,0,1)
		= p_{\mathrm{Q}}(1,1,0)
		= \frac{1}{4},\\
		p_{\mathrm{Q}}(a,b,c) = 0
		\quad\text{for all other }(a,b,c),
	\end{gathered}
	\label{eq:quantum_probability}
\end{equation}
so that all nonzero events satisfy $a\oplus b\oplus c = 0$, thus achieving the target parity constraint impossible for any classical model $p_{\mathrm{cl}}(a,b,c)$.

Within this known physical framework, i.e., the triangular network structure described above, our coherence criterion kernel, which corresponds to Eq.~(1) in the main text where $n=1$, can be rewritten as:
\begin{equation}
	\mathcal{Q} = \min_{\mathcal{E}_\text{cl}} \sum_{i=0}^{7}
	\left | P_{\mathcal{E}_{\text{V}}} (i \mid \rho_{\text{out}} )
	- P_{\mathcal{E}_\text{cl}} (i \mid \rho^{(\text{d})}_{\text{out}}) \right|,
\end{equation}
where the checkpoint operation $\mathcal{E}_{\text{V}}$ at each node is a BS transformation and $\mathcal{E}_{\mathcal{I}}=\mathcal{E}_\text{cl}$ is prescribed according to the known classical realistic model [Eq.~(\ref{eq:classical_model})], which is incapable of creating coherence. This also corresponds to the kernel of Eq.~(2) in the main text. Here, the population measurement is defined with respect to the postselected binary-outcome probability distribution $p(a,b,c)$ associated with the three parties, so that the label $i$ simply enumerates the eight possible triples $(a,b,c)\in\{0,1\}^3$.

The state $\rho_{\text{out}}$ corresponds to the coherent path superposition state from Eq.~\eqref{eq:path_superposition_initial_state}, and, as highlighted above, the process $\mathcal{E}_{\text{V}}$ represents the overall operation of the three 50:50 BSs in the triangular network. Consequently, $P_{\mathcal{E}_{\text{V}}} \left(i \mid \rho_{\text{out}} \right)$, i.e., $P_{\mathcal{E}_{\text{V}}} \left(i \mid \rho_{\text{out}} \right)=P_{\mathcal{E}_{\text{V}}} \left(a,b,c\mid \rho_{\text{out}} \right)$ in Eq.~(2) in the main text, reproduces the quantum output distribution $p_{\mathrm{Q}}(a,b,c)$ obtained from Eq.~\eqref{eq:quantum_probability}.

The incapable process $\mathcal{E}_\text{cl}$, which lacks the ability to create quantum coherence, on the other hand, is restricted to classical triangle network models of the form given in Eq.~\eqref{eq:classical_model}. In our general discussion, the diagonal state $\rho^{(\text{d})}_{\text{out}}$ would normally be constructed via an additional experimental configuration. In the present structured scenario, however, the process $\mathcal{E}_\text{cl}$, as described by the classical model of Eq.~\eqref{eq:classical_model}, directly yields the corresponding population probabilities $p_{\mathrm{cl}}(a,b,c)$, thereby eliminating the need for another experimental configuration to construct $\rho^{(\text{d})}_{\text{out}}$. Moreover, $\rho^{(\text{d})}_{\text{out}}$ retains its usual physical meaning as the fully dephased version of $\rho_{\text{out}}$, and the population distribution associated with $\rho^{(\text{d})}_{\text{out}}$ in this scenario it naturally falls within the set described by the classical model of Eq.~\eqref{eq:classical_model}.

Thus, according to the above discussions, the minimization over incapable processes $\mathcal{E}_\text{cl}$ is effectively reduced to an optimization over the classical network parameters $\gamma$, $p_X$ and $q_X$, and the criterion kernel becomes:
\begin{equation}
	\mathcal{Q}
	= \min_{\gamma,{p_X,q_X}}
	\sum_{(a,b,c)\in\{0,1\}^3}
	\bigl|
	p_{\mathrm{Q}}(a,b,c) - p_{\mathrm{cl}}(a,b,c)
	\bigr|.
\end{equation}
By inserting the explicit quantum probabilities from Eq.~\eqref{eq:quantum_probability}, we obtain the concrete form:
\begin{widetext}
\begin{eqnarray}
\mathcal{Q}&=&\!\!\! \min_{\gamma, p_X, q_X} (\left| \tfrac{1}{4} \!-\!p_{\mathrm{cl}}(000) \right| \!+ \! \left| \tfrac{1}{4}  \!- \! p_{\mathrm{cl}}(011) \right|  \!+ \! \left| \tfrac{1}{4} \! - \! p_{\mathrm{cl}}(101) \right| \!+ \! \left| \tfrac{1}{4}  \!- \! p_{\mathrm{cl}}(110) \right| \!\nonumber\\
&&\ \ \ \ \ \ \ \ \ \ + \! \left| 0  \!- \! p_{\mathrm{cl}}(001) \right| \!+ \! \left| 0  \!- \! p_{\mathrm{cl}}(010) \right|  \!+ \! \left| 0  \!- \! p_{\mathrm{cl}}(100) \right| \!+ \! \left| 0 \! - \! p_{\mathrm{cl}}(111) \right|) > 0.
\end{eqnarray}
\end{widetext}
This implies the criterion for coherence transfer in a triangular network, as given in the main text. Ref.~\cite{Bibak24} has already shown that the quantum distribution $p_{\mathrm{Q}}(a,b,c)$ cannot be reproduced by any classical triangle network under the classical realistic model, Eq.~\eqref{eq:classical_model}. As a result, no choice of $\gamma$, $p_X$, and $q_X$ can simultaneously make all eight absolute value terms vanish, and the minimized value of our criterion kernel necessarily satisfies $\mathcal{Q} > 0$. Following the same line of analysis, our criterion can also be applied to the $n$-party network defined in Ref.~\cite{Bibak24}.

In summary, this example shows that, under known structured network assumptions, the coherence criterion detects the nonclassical coherence present in the triangular network. In particular, once the network's internal architecture and the set of processes that cannot create quantum coherence are specified, our criterion can be evaluated using a single coherence state configuration ($n=1$), without changing the state or modifying the local measurement setting, in direct analogy with the scenario analyzed in Ref.~\cite{Bibak24}. Moreover, coherence criteria (1) and (2) in the main text not only detect the nonclassical coherence in the triangular network but also confirm that the BSs' operation can create coherence. Overall, this shows that the same formalism we employ to certify coherence transfer in general quantum networks also naturally applies to verifying coherence resources in networks whose internal structure is known.

\section{The general number of input and output state settings: $n\geq 2$} \label{app:number of state}

It is important to note that when there is only a single input and output state setting ($n = 1$) in criterion~(1) in the main text, the criterion kernel simplifies to the following form:
\begin{equation}
	\mathcal{Q} = \min_{\mathcal{E}_\mathcal{I}} \sum_{i=0}^{d-1}
	\left | P_{\mathcal{E}_{\text{V}}} (i\mid \rho_{\text{out},1} )
	- P_{\mathcal{E}_\mathcal{I}}(i \mid \rho^{(\text{d})}_{\text{out},1}) \right|.
	\label{eq:single_output_criterion}
\end{equation}
This expression represents the search for an incapable process $\mathcal{E}_\mathcal{I}$~\cite{Kuo19} that maps the diagonal state $\rho^{(\text{d})}_{\text{out},1}$ to a unique probability distribution identical to the population distribution of $\mathcal{E}_{\text{V}}(\rho_{\text{out},1})$, i.e., $P_{\mathcal{E}_{\text{V}}}(i\mid \rho_{\text{out},1})$ for $i = 0, 1, \dots, d-1$. Such a transformation lies within the set of operations defined in the theory of quantum process capability~\cite{Kuo19} as incapable of creating coherence, namely, those for which any incoherent input state remains incoherent under the operation. Consequently, there always exists an incoherent operation $\mathcal{E}_\mathcal{I}$ that evolves $\rho^{(\text{d})}_{\text{out},1}$ to match the unique population distribution of $\mathcal{E}_{\text{V}}(\rho_{\text{out},1})$, yielding a vanishing criterion kernel, i.e., $\mathcal{Q} = 0$.

From another perspective, when there is only a single input and output state setting ($n=1$), the number of independent equations is insufficient relative to the number of unknowns in the kernel of criterion~\eqref{eq:single_output_criterion}. Specifically, there are $d^2$ unknowns---namely the terms $\chi_{\mathcal{I},in,in}$ in $P_{\mathcal{E}_\mathcal{I}}(i|\rho^{(\text{d})}_{\text{out},k})=\sum_{n=0}^{d-1} \rho_{k,nn} \chi_{\mathcal{I},in,in}$, where $i,n = 0,1,\ldots,d-1$. In contrast, only $d$ independent equations are available, corresponding to the $d$ equations obtained by summing over $i$ in Eq.~\eqref{eq:single_output_criterion}. According to basic principles of linear algebra, such an underdetermined system always admits at least one solution that satisfies $Q = 0$, even under the additional restriction that the processes are incapable of creating coherence.

Consequently, detecting coherence transfer requires $d$ distinct input and output state settings to fully characterize the process. This requirement arises because, in such cases, there are $d^2$ independent equations and $d^2$ unknowns; when both the network process and the checkpoint operation are capable, only a single non-zero optimal solution exists. In special cases, the number of unknowns---the terms $\chi_{\mathcal{I},in,in}$ in $P_{\mathcal{E}_\mathcal{I}}(i|\rho^{(\text{d})}_{\text{out},k})$ as discussed above---can be effectively reduced if some of the chosen output states $\rho^{(\text{d})}_{\text{out},k}$ have vanishing diagonal elements. This reduction in the number of variables allows for the detection of coherence transfer even when the number of input and output state settings satisfies $2 \leq n < d$.

\section{Coherence transfer criterion for coherence dynamics}

\subsection{Proof of the dynamics criterion}

To prove the dynamics criterion for creating coherence, Eq.~(3) in the main text, we review the following steps for the criterion kernel:
\begin{equation}
\mathcal{Q}_{t}:=\min_{\mathcal{E}_{\mathcal{I}}}\sum_{k=1}^{n}\!\sum_{i=0}^{d-1}\left|P(i|\rho_{k,t})-P_{\mathcal{E}_{\mathcal{I}}}(i|\rho_{k,t_0}^{(\text{d})})\right|.\label{criterion2}
\end{equation}
For an unknown dynamics $\mathcal{E}_{t,0}$ in an open quantum system coupled to a reservoir~\cite{Breuer16}, we initially consider $n$ states with different initial conditions at time 0, denoted by $\rho_{k,0}$, for $k=1,2,..., n$. For each $k$, we perform population measurements at time $t_0$ on the states $\rho_{k,t_0}$ that have evolved under the dynamics $\mathcal{E}_{t_0,0}$ from 0 to $t_0$. Subsequently, we also conduct population measurements at a later time $t$, where $t > t_0$, on the states $\rho_{k,t}$ that have evolved under the dynamics $\mathcal{E}_{t,0}$ from 0 to $t$.

If the system and the reservoir are coupled under the Markovian approximation and $\mathcal{E}_{t,0}$ is incapable of forming coherence between the basis states in $\{\ket{i}\}$ over time, then the system states are incoherent with diagonal density matrices: 
\begin{eqnarray}
\rho_{k,t_0}&=&\rho_{k,t_0}^{(\text{d})},\ \ \rho_{k,t}=\rho_{k,t}^{(\text{d})},
\end{eqnarray}
and the evolution of $\rho_{k,t_0}$ to $\rho_{k,t}$ undergoes Markovian dynamics which can be described by a completely positive (CP) quantum operation. The entire dynamics $\mathcal{E}_{t,0}$ is called CP-divisible~\cite{Breuer16} with the following decomposition:
\begin{eqnarray}
\mathcal{E}_{t,0}=\mathcal{E}_{t,t_0}\circ\mathcal{E}_{t_0,0},
\end{eqnarray}
 where the dynamics $\mathcal{E}_{t,t_0}$ describes the additional evolution from $t_0$ to $t$. It is worth noting that $\mathcal{E}_{t,t_0}$ corresponds to the role of $\mathcal{E}_{\text{V}}$ in Eq.~(1) in the main text, and it is an incapable process of coherence creation. Therefore, we can always find an incapable process $\mathcal{E}_{\mathcal{I}}$ such that the value of the criterion kernel $\mathcal{Q}_{t}$ is zero. When experimental results meet criterion~(3), $\mathcal{E}_{t,0}$ is confirmed to effectively generate coherence between states over time in the open quantum system coupled to the reservoir Markovianly.

\subsection{Example of the dynamics criterion: quantum transport through a double quantum dot}

Experimentally distinguishing quantum from classical transport in nanostructures remains a key challenge for understanding transport phenomena and developing quantum electronic devices~\cite{Lambert10}. As mentioned in the introduction in the main text, using time-domain methods to verify quantum coherence, such as testing the Leggett-Garg inequality, can be very demanding. We show here that our criterion~(3) in the main text holds under invasive population measurements by modeling single-electron transport through a double quantum dot. Compared with the time periods identified by the Leggett-Garg-type approach~\cite{Lambert10}, the criterion $\mathcal{Q}_{t}>0$ (Fig.~3 in the main text) can detect a much larger window of quantum coherence.

For the setup in Fig.~1(d) of the main text, we assume that the double quantum dot (DQD) is weakly coupled to the leads at a large bias. Its Hamiltonian is
\begin{equation}
H_{\mathrm{DQD}}=\Delta(\left|L\right\rangle\!\!\left\langle R\right|+\left|R\right\rangle\!\!\left\langle L\right|),\label{h}
\end{equation}
with the electron state basis
$\{\left|L\right\rangle\!,\!\left|R\right\rangle\!,\!\left|0\right\rangle\}$ where $\Delta$ is the tunneling amplitude between the left-dot
and right-dot electron states $\left|L\right\rangle,\left|R\right\rangle$ and $\left|0\right\rangle$ is the state of the electron absent. The transport between dots and leads is described by the self-energy
\begin{equation}
\Sigma[\rho]=-\frac{1}{2}\sum_{\alpha=L,R}\Gamma_{\alpha}[s_{\alpha}s_{\alpha}^{\dag}\rho-2s_{\alpha}^{\dag}\rho
s_{\alpha}+\rho s_{\alpha}s_{\alpha}^{\dag}],\label{self}
\end{equation}
where $\rho$ is the electron's density operator, $s_{L}=\left|0\right\rangle\!\!\left\langle L\right|$, $s_{R}=\left|R\right\rangle\!\!\left\langle 0\right|$, and $\Gamma_{L}$ and $\Gamma_{R}$ are the left and right tunneling rates, respectively. With Eqs.~(\ref{h}) and~(\ref{self}) under weak coupling to the contacts, we use a generalized master equation description of the electron transport~\cite{Brandes05}. We also assume that charge detectors, such as those used with a quantum point contact~\cite{Fujisawa06}, are used for the population measurements. According to criterion~(3) in the main text, we define $\left|0\right\rangle$, $\left|L\right\rangle$, and $\left|R\right\rangle$ as $\left|0\right\rangle$, $\left|1\right\rangle$, and $\left|2\right\rangle$, respectively. The probabilities $P(i|\rho_{k,t})$ and $P_{\mathcal{E}_{\mathcal{I}}}(i|\rho_{k,t_0}^{(\text{d})})$ can then be derived from the master equation and the incapable processes $\mathcal{E}_{\mathcal{I}}$~\cite{Kuo19}, respectively. The subsequent minimization of $\mathcal{Q}_t$ is performed via semidefinite programming (SDP)~\cite{Lofberg2004Yalmip,Toh1999Sdpt3,Andersen2000Mosek}. (See Sec.~\ref{sdp}4 for a detailed discussion of the minimization over all possible incapable processes.) For the setting $\Gamma_{L}=4$, $\Gamma_{R}=0.1$, and $\Delta=1$, the non-vanishing $\mathcal{Q}_{t}$ indicates regions of coherence formation in quantum transport, as shown in Fig.~3 in the main text. Compared with the most effective time-domain witness~\cite{Li12,Kofler13}, our result is consistent with the one reported in Ref.~\cite{Li12}. However, the present criterion is more practical because it does not require additional reliable reference states or propagator measurements. Moreover, the role of $\mathcal{E}_{\text{V}}$ is played by the additional system dynamics from $t_0$ to $t$. This makes detection of coherence dynamics experimentally feasible using only population measurements.

\begin{figure*}[t]
\includegraphics[width=14.5cm]{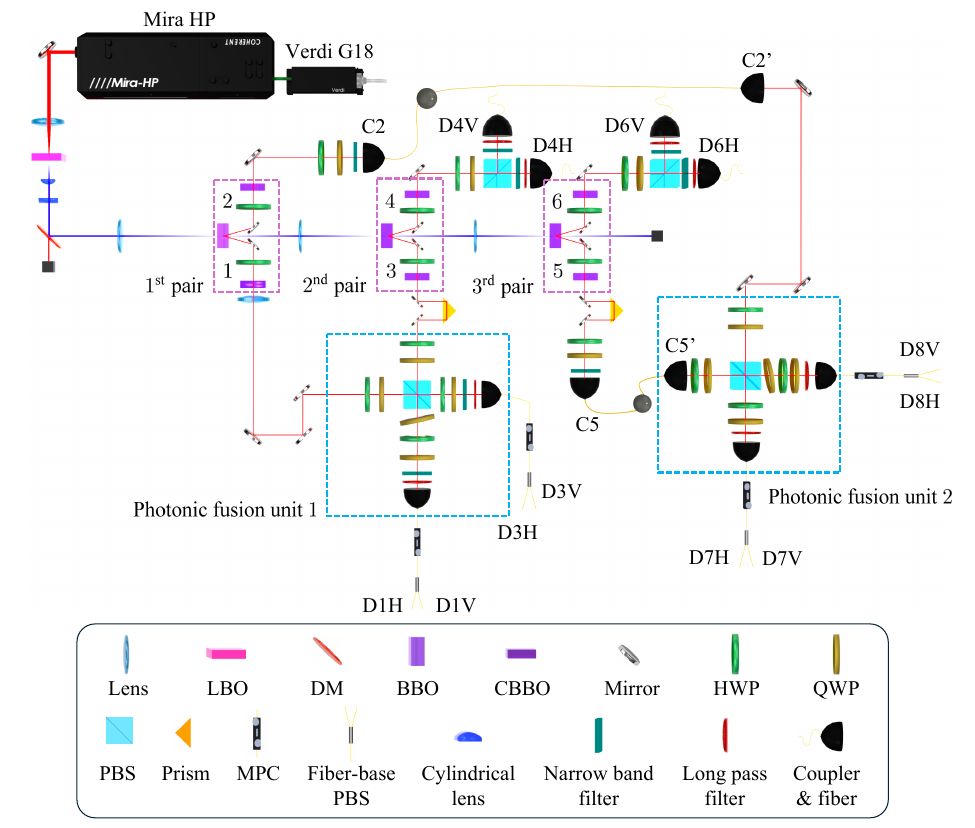}
\caption{Full experimental setup for coherence transfer and detection in multi-photon entanglement networks. Abbreviations: $10$ cm focus lens (Lens), LiB$_3$O$_5$ (LBO), dichroic mirrors (DM), $\beta$-barium borate (BBO), compensating $\beta$-barium borate (CBBO), half-wave plate (HWP), quarter-wave plate (QWP), polarizing beam splitter (PBS), motorized fiber polarization controller (MPC). The abbreviations Ck, DkH, and DkV denote the No.$k$ coupler for photons in the $k$th spatial mode and the corresponding detectors for the $H$ and $V$ measurement outcomes, respectively.}
\label{setup}
\end{figure*}

\begin{table*}
\renewcommand\arraystretch{1.5}
\caption{\label{tab:table3}Summary of the three SPDC entangled photon pairs created and their related qualities. Here, the fidelities of the four-photon Greenberger-Horne-Zeilinger (GHZ)-type states~\cite{GHZ07} are measured by comparing the target GHZ-type states to the experimentally produced states using the first photonic fusion unit on the first and second pairs, and with respect to the experimental states via the second photonic fusion unit on the first and third pairs, respectively.}
\begin{ruledtabular}
\begin{tabular}{cccc}
 &{\rm Pair 2}&{\rm Pair 1}&{\rm Pair 3}\\ \hline
 {\rm Two-fold coincidence rate (Hz)}&$80 \times 10^3$&$76 \times 10^3$ &$38 \times 10^3$ \\
{\rm Pump power (W)}&$ 1.24$&$ 1.07$&$ 0.89$\\
  {\rm State fidelity}&$(93.03\pm 0.09)\%$&$(94.02\pm 0.09)\%$ &$(93.05\pm 0.13)\%$\\
\multirow{2}{*}{{\rm Four-fold coincidence rate (Hz)}}&\multicolumn{2}{c}{$\sim 42$}&$-$\\
 &$-$&\multicolumn{2}{c}{$\sim 27$}\\
 \multirow{2}{*}{{\rm GHZ type state fidelity}}&\multicolumn{2}{c}{$(74.77\pm 0.36)\%$}&$-$\\
 &$-$&\multicolumn{2}{c}{$(71.51\pm 0.37)\%$}\\
  \multirow{2}{*}{{\rm Fusion process fidelity}}&\multicolumn{2}{c}{$(81.55\pm 0.76)\%$}&$-$\\
 &$-$&\multicolumn{2}{c}{$(75.231\pm 0.24)\%$}\\
% Fusion process fidelity & & & 49
\end{tabular}
\end{ruledtabular}\label{3pairdata}
%\footnote{The measured four-photon GHZ state fidelity by QST.}
%\footnotemark[2]{The measured four-photon GHZ state fidelity by minimum five observable.}
\end{table*}

\section{Properties of the proposed criterion kernel $\mathcal{Q}$ and criteria~Eqs.~(1) and~(3)}

We discuss several properties of the proposed criterion kernel $\mathcal{Q}$ and the corresponding criterion. Because Eq.~(1) and its temporal analog Eq.~(3) in the main text are computed using SDP~\cite{Lofberg2004Yalmip,Toh1999Sdpt3,Andersen2000Mosek}, these properties are primarily tied to the SDP.

1.~Tightness of the criterion kernel. The tightness of $\mathcal{Q}$ is defined with respect to a given checkpoint operation $\mathcal{E}_{\text{V}}$. Although an uncharacterized choice of $\mathcal{E}_{\text{V}}$ may lead to false negatives ($\mathcal{Q} = 0$ for a capable process), the mathematical bound itself remains strictly tight. That is, a nonzero kernel $\mathcal{Q}>0$ guarantees that the network process can transfer coherence and that $\mathcal{E}_{\text{V}}$ can create coherence.

2.~Optimality guaranteed via semidefinite programming. The evaluation of $\mathcal{Q}$ is formulated as a standard SDP, which mathematically guarantees optimality. In convex optimization, such as SDP, any local minimum is the global minimum. Formulating $\mathcal{Q}$ as an SDP ensures that the algorithm computes the true lower bound rather than getting trapped in local extrema.

3.~Detection power and performance comparison. Detection power can be defined by comparing criteria based on the physical importance of the quantities of interest. Here, as already illustrated in the temporal case, the detection power of $\mathcal{Q}_t>0$ is demonstrated by detecting coherence transfer in single-electron transport through a double quantum dot. See Fig.~3 in the main text. Compared with the time periods identified by violations of the one-initial-state, three-time-point Leggett-Garg-type inequality~\cite{Lambert10} (which a classical Markovian system under noninvasive charge measurement satisfies), our criterion can detect a much broader window of quantum coherence, given three initial states ($n=3$) for considering $\mathcal{Q}$ and implementing minimization via SDP, and within the same limits to which the master equation formalism applies. This identifies coherence transfer through the coexistence of static and dynamic coherence, distinct from approaches that rule out a classical description of transport through a nanostructure using a Leggett-Garg-type inequality.

4.~SDP complexity. To evaluate the computational complexity of obtaining $\mathcal{Q}_t$, we solved the optimization using the MOSEK solver in a conic/SDP framework. After the presolve phase, the effective problem size for detecting one-qubit (two-qubit) coherence transfer involves 24 (256) constraints, 9 (65) scalar variables, and 7 (17) matrix variables, equivalent to 96 (1,104) scalarized variables. The solver employed the interior-point method. The computational complexity per iteration was dominated by the matrix factorization, requiring $1.03 \times 10^4$ ($1.01 \times 10^7$) floating-point operations. Due to the high sparsity of our mathematical formulation, the non-zero entry count before and after factorization remains well-controlled (e.g., at $3.29 \times 10^4$ for two-qubit systems), effectively preserving sparse linear algebra performance without triggering the numerical instability typical of dense SDP solvers. The algorithm converges quickly and tightly, reaching the exact global optimum in only 7 iterations under a strict primal/dual feasibility tolerance of $10^{-8}$. Furthermore, because the optimization at each time point is solved independently, the total computational cost scales strictly linearly with the number of evaluated time points, confirming that the temporal extension remains lightweight and scalable.

5.~Error and its propagation in a measured $\mathcal{Q}$. The error and its propagation in the measured kernel $\mathcal{Q}$, shown in Fig.~4 in the main text and Fig.~S8 later, are first derived from the fact that photon coincidence counts follow a Poisson distribution. We then calculate the measured probability error by propagating the counting noise. For a measured kernel, $\mathcal{Q}=\!\sum_{k=1}^{n}\sum_{i=0}^{d-1}\!\left|P_{\mathcal{E}_\text{V}}(i|\rho_{\text{out},k})\!-\!P_{\mathcal{E}_{\mathcal{I}_\text{min}}}(i|\rho_{\text{out},k}^{(\text{d})})\right|$, where $\mathcal{E}_{\mathcal{I}_\text{min}}$ is the incapable process for coherence creation that minimizes $\mathcal{Q}$, it can be represented as
\begin{equation}
\mathcal{Q}=\sum_{k=1}^{n}\sum_{i=0}^{d-1}c_{k,i}\left[P_{\mathcal{E}_\text{V}}(i|\rho_{\text{out},k})-P_{\mathcal{E}_{\mathcal{I}_\text{min}}}(i|\rho_{\text{out},k}^{(\text{d})})\right],\label{newq}
\end{equation}
where $c_{k,i}\in\{+1,-1\}$ is chosen so that the difference between $P_{\mathcal{E}_\text{V}}(i|\rho_{\text{out},k})$ and $P_{\mathcal{E}_{\mathcal{I}_\text{min}}}(i|\rho_{\text{out},k}^{(\text{d})})$ is nonnegative. Both $P_{\mathcal{E}_\text{V}}(i|\rho_{\text{out},k})$ and $\rho_{\text{out},k}^{(\text{d})}$ are derived from the measured photon coincidence counts. Moreover, $\mathcal{E}_{\mathcal{I}_\text{min}}$ is known and obtained via SDP. Therefore, the variance and standard deviation of $\mathcal{Q}$~(\ref{newq}) can also be determined using the error-propagation formula for independent variables under Poissonian photon-counting statistics.

6.~SDP numerical errors. As described in the fourth point above, due to the formulation's high sparsity, the number of nonzeros remains the same before and after factorization at $3.29 \times 10^4$, effectively scaling without triggering the notorious numerical errors typical of SDP solvers. The algorithm converges quickly, reaching the optimal solution in only 7 iterations with a primal/dual feasibility tolerance of $10^{-8}$.

\section{Entangled photon sources and multi-photon entanglement}

\subsection{Laser system and pumping source}

Figure~\ref{setup} shows the complete schematic of our laser system and multi-photon experimental setup. See also Fig.~4(a) in the main text. First, we use a diode-pumped solid-state CW laser, the Verdi-G18, to pump a high-power ultrafast Ti:sapphire oscillator, the Mira HP. An ultrafast infrared pulsed laser beam is generated with an output power of about $2.95$~W when the Verdi-G18 is operated at $13.3$~W. The ultrafast infrared pulsed laser is focused onto the up-converted crystal $\rm{LiB_{3}O_{5}}$ (LBO) to produce ultraviolet pulses via up-conversion. The ultraviolet laser beam is circularized using horizontal and vertical cylindrical lenses. Then, seven dichroic mirrors (DMs) filter out infrared light and transmit ultraviolet light. The resulting ultraviolet laser beam has a center wavelength of $390$~nm, a pulse duration of $144$~fs, a repetition rate of $76$~MHz, and an average power of $1.0$~W.

\subsection{Entangled photon pairs}

The ultraviolet pulse laser beam is focused onto three $\beta\rm{-BaB_2O_4}$ (BBO) crystals using focusing lenses with a $10$~cm focal length to generate entangled photons via spontaneous parametric down-conversion (SPDC)~\cite{Kwiat95,Pan12} with high brightness, including compensation for the walk-off effect using an additional correction BBO (CBBO) crystal. The pump powers for the first, second, and third BBOs are $1.24$~W, $1.07$~W, and $0.89$~W, respectively. After collection with fiber couplers, we use silicon avalanche photodiode single-photon detector modules to measure the number of photons each detector receives, and a coincidence logic processing unit (a field-programmable gate array-based coincidence unit) to measure the coincidence counts. In front of each fiber coupler, a narrow-band filter (central wavelength $780$~nm, full width at half maximum (FWHM) $3$~nm, transmission rate $\ge99\%$) increases the spectral indistinguishability of the photons, facilitating the observation of the Hong-Ou-Mandel-type effect. Additionally, a long-pass filter (transmitting from $460$~nm to $1200$~nm with a transmission rate $\ge80\%$) eliminates scattered light from the ultraviolet pump. A half-wave plate (HWP) and a quarter-wave plate (QWP) are used for polarization analysis. A QWP and a polarizer (Pol.) are combined for projective measurements. The relative phase between the two-photon amplitudes in a photonic fusion unit is adjusted by rotating an additional QWP around its vertically oriented fast axis. 

These configurations and settings produce three pairs of SPDC-entangled photons corresponding to the target state $\left|\psi^-\right\rangle=(1/\sqrt{2})(\left|H\right\rangle\otimes\ket{V}-\left|V\right\rangle\otimes\ket{H})$, where $\ket{H}$ and $\ket{V}$ represent the horizontal and vertical polarization states, respectively. The photon pair produced in modes $1$-$2$ is the first pair, while the pairs created in modes $3$-$4$ and $5$-$6$ are the second and third pairs, respectively. See Fig.~\ref{setup}. To assess how well the prepared state matches the target state, we use a set of measurement wave plates, comprising a QWP and an HWP, to determine the state fidelity. The state fidelity is defined as $F_{s} = {\rm tr}(\rho_{\rm expt}\ket{\psi^-}\!\!\bra{\psi^-})$, where $\rho_{\rm expt}$ represents the experimental two-photon state. We calculate the state fidelity by measuring three expectation values of observables: $\langle{X}{X}\rangle$, $\langle{Y}{Y}\rangle$, and $\langle{Z}{Z}\rangle$, where $\langle{M_1}{M_2}\rangle$ = ${\rm tr}(\rho_{\rm expt} M_1\otimes M_2)$, and $M_{1,2}\in\{X, Y, Z\}$ are the Pauli operators. Experimentally, these operators are defined as: $X=\ketbra{+}{+}-\ketbra{-}{-}$, $Y=\ketbra{R}{R}-\ketbra{L}{L}$, and $Z=\ketbra{H}{H}-\ketbra{V}{V}$, where $\ket{+}=1/\sqrt{2}(\ket{H}+\ket{V})$, $\ket{-}=1/\sqrt{2}(\ket{H}-\ket{V})$, $\ket{R}=1/\sqrt{2}(\ket{H}+i\ket{V})$, and $\ket{L}=1/\sqrt{2}(\ket{H}-i\ket{V})$.

We detect $80\times10^3$ photon pairs per second from the first BBO and about $76\times10^3$ pairs per second from the second BBO. The fidelity of the first entangled photon pair is $(93.03\pm0.09)\%$, while that of the second pair is $(94.02\pm0.09)\%$. The third pair of SPDC entangled photons in modes $5$-$6$ is generated by directing the $390$~nm pump beam through the first two BBOs, then refocusing it onto the third BBO, as shown in Fig.~\ref{setup}. The pump power for the third BBO is $0.89$~W. The resulting state fidelity is $(93.05\pm0.13)\%$ relative to the target state $\ket{\psi^-}$, with a creation rate of $38\times10^3$ pairs per second.

\subsection{Multi-photon entanglement sources}

After creating all three pairs, we route the photons in modes $1$ and $3$ to photonic fusion unit 1 and the photons in the original modes $2$ and $5$ to photonic fusion unit 2 via single-mode optical fibers. Due to space limitations on our optical table, we use fiber-based photon transmission to fuse photons in photonic fusion unit 2, with each fiber 2 meters long~\cite{Im21}. We align the photons in modes $2'$ and $5'$, corresponding to couplers C2' and C5', respectively, before the polarizing beam splitter (PBS) to ensure they are parallel, thereby maintaining consistent beam divergence and convergence. Additionally, as shown in Fig.~\ref{setup}, to compensate for the unequal lengths of the free-space paths of the photons from the first and third BBOs, three mirrors are placed in mode $2'$ before the PBS to make the beam sizes and path lengths as similar as possible.

Then, we optimize the coupling efficiencies for D7 and D8 using entangled photons from modes $2'$ and $5'$, respectively, to ensure spatial overlap in the interferometer. However, because the photons are coupled twice through the two optical fibers, the yield of the two pairs of entangled photons decreases, and the coupling efficiency drops accordingly.

Table~\ref{3pairdata} summarizes the measured two- and four-photon coincidence count rates (photon-pair creation rates), as well as the state and process fidelities for photonic fusion at the respective pump powers. The four-photon states are generated via photonic fusion to approximate a GHZ-type target state. Specifically, the first photonic fusion process produced the state $\ket{G_4} = (1/\sqrt{2}) \left( \ket{H_{1} V_{2} H_{3} V_{4}} + \ket{V_{1} H_{2} V_{3} H_{4}} \right)$, where the subscripts denote the mode numbers, whereas the second photonic fusion process produced the state $\ket{G_4} = (1/\sqrt{2})\left( \ket{H_{1} V_{2} V_{5} H_{6}} + \ket{V_{1} H_{2} H_{5} V_{6}} \right)$. These four-photon state fidelities reflect the performance of coherence transfer in the multi-photon entangled networks we have created. To verify that the source indeed generates genuine six-photon entanglement, we evaluate the entanglement witness associated with the six-qubit GHZ state $\Ket{G_6} = (1/\sqrt{2})\left( \Ket{H_1 V_2 H_3 V_4 V_5 H_6} + \Ket{V_1 H_2 V_3 H_4 H_5 V_6} \right)$. If the witness value is negative, specifically when $\mathrm{tr}(\mathcal{W}_{\text{GHZ}_6} \rho_{G_6}) < 0$, it confirms the presence of genuine multipartite entanglement. The witness~\cite{Toth05} is defined as:
\begin{equation}
	\mathcal{W}_{\text{GHZ}_6} := 3I - 2 \left[ \frac{S_1^{\text{GHZ}_6} + I}{2} + \prod_{k=2}^{6} \frac{S_k^{\text{GHZ}_6} + I}{2} \right],\label{ew}
\end{equation}
where
\begin{equation}
	\begin{aligned}
		S_1^{\text{GHZ}_6} &= \prod_{k=1}^{6} \sigma_X^k, \\
		S_k^{\text{GHZ}_6} &= \sigma_Z^{k-1} \sigma_Z^k, \quad \text{for } k = 2, 3, \ldots, 6,
	\end{aligned}
\end{equation}
where $\sigma_X$ and $\sigma_Z$ denote the Pauli $X$ and $Z$ operators, respectively. In this approach, only two observables, $\langle X^{\otimes 6} \rangle$ and $\langle Z^{\otimes 6} \rangle$, are needed to evaluate the entanglement witness. The measured expectation values are $\langle X^{\otimes 6} \rangle = 0.4539 \pm 0.0365$ and $\langle Z^{\otimes 6} \rangle = 0.7355 \pm 0.0275$. The average six-fold coincidence rate is approximately $150$ photons over $3$ hours, corresponding to $0.014$ counts per second. Using these expectation values, we calculate the entanglement witness for the six-photon GHZ state as $\text{tr}(\mathcal{W}_{\text{GHZ}_6} \rho_{G_6}) = -0.0572 \pm 0.0488$. This confirms the presence of genuine multipartite entanglement and supports the effectiveness of the multi-photon entangled networks we created for coherence transfer. See Sec.~\ref{cen} for details on the relationship between the coherence transfer criterion and the genuine multipartite entanglement witness.

\section{Detection of one-qubit coherence transfer in multi-photon entanglement networks}

We use remote state preparation (RSP)~\cite{Pati00,Bennett01,Chen24}, photonic fusion~\cite{Pan98,Pan01,Pan03,Lu07}, and entangled photon pairs to implement the three stages of measurement-based coherence transfer: input coherence, evolution, and output, in our four- and six-photon entanglement quantum networks. The fundamental principles are outlined below. Using these elements, we will explain how to transfer and detect one-qubit coherence in multi-photon entanglement networks.

\begin{figure}[t]
	\includegraphics[width=6.5cm]{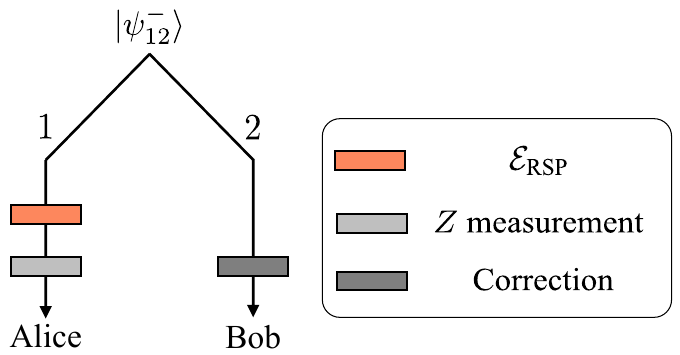}
	\caption{Remote state preparation (RSP). With the shared pairs, $\ket{\psi^-}_{12}$, where the photon in spatial mode $1$ is sent to Alice and the photon in mode $2$ is sent to Bob, Alice aims to help Bob prepare a state $\ket{s}$ or $\ket{s^{\perp}}$ for Bob. For example, the state $\ket{s}$ can be a state with coherence expressed as follows: $\ket{s}=(\ket{0}+e^{i\phi}\ket{1})/\sqrt{2}$, where $\{\ket{i}\}=\{\ket{0},\ket{1}\}$ is an orthonormal basis, called the laboratory basis in our experiment (also called the computational basis). A measurement performed on a laboratory basis is called the $Z$ measurement. To implement the RSP protocol, Alice first performs the prescribed measurement and sends the result to Bob (not shown). Here, this measurement is performed using the quantum operation $U^{\dag}$, denoted as $\mathcal{E}_{\text{RSP}}$, followed by a $Z$ measurement. Then, Bob receives $\ket{s}$ or $\ket{s^{\perp}}$ of his qubit, which depends on Alice's measurement result, either $\ket{s^{\perp}}$ or $\ket{s}$. Here, the RSP scenario is implemented for Bob's prepared state orthogonal to Alice's measurement result, without Bob's corrections, compared to the original protocol.}
	\label{fig:rsp}
\end{figure}

\subsection{Remote state preparation}

The state $\ket{\psi^{-}}$ has the following symmetry property:
\begin{equation}
\ket{\psi^{-}}=\frac{1}{\sqrt{2}}(U\ket{s_{0}} \otimes U\ket{s_{0}^{\perp}}-U\ket{s_{0}^{\perp}} \otimes U\ket{s_{0}}),
\end{equation}
where $\ket{s_{0}}$ and $\ket{s_{0}^{\perp}}$ are arbitrary orthonormal states, and $U$ is a single-qubit unitary operator. Therefore, it is clear that, after measuring the basis $\{U\ket{s_{0}},U\ket{s_{0}^{\perp}}\}$ on one qubit of the entangled pair, the other qubit becomes: $\ket{s}=U\ket{s_0}$ or $\ket{s^{\perp}}=U\ket{s^{\perp}_0}$. For Alice, the information sender, she can utilize this simple feature of $\ket{\psi^{-}}$ to prepare a state, either $\ket{s}$ or $\ket{s^{\perp}}$, for Bob, the remote information receiver, depending on her measurement result, which is either $\ket{s^{\perp}}$ or $\ket{s}$. Compared to the original protocol, we consider a scenario in which, in the current RSP protocol, Bob's prepared state is orthogonal to Alice's measurement result, rather than Bob needing to make corrections to ensure his result state matches Alice's state. See Fig.~\ref{fig:rsp}, where the photon in spatial mode $1$ is sent to Alice and the photon in mode $2$ is sent to Bob.

It is worth noting that Alice's measurement capability includes her ability to implement a unitary operation, $U^\dag$, related to the operation $U$ such that $U\ket{s_{0}}$ or $U\ket{s_{0}^{\perp}}$ becomes the target superposition state. This can be observed from the following main experimental steps used to implement the RSP protocol.

First, Alice performs the operation $U^{\dag}$ so that $\ket{\psi^{-}}$ becomes
\begin{equation}
U^\dag\ket{\psi^{-}}=\frac{1}{\sqrt{2}}(\ket{s_{0}}\otimes U\ket{s_{0}^{\bot}}-\ket{s_{0}^{\bot}}\otimes U\ket{s_{0}}).\label{explain1}
\end{equation}
In our experimental RSP, $U^\dag$ is implemented using a setup with a QWP and an HWP, denoted as $\mathcal{E}_{\text{RSP}}$.

Second, Alice's qubit is measured in the experimental basis $\{\ket{i}\}=\{\ket{s_{0}},\ket{s_{0}^{\bot}}\}=\{\ket{0},\ket{1}\}$. Experimentally, this is achieved using Alice's polarization analyzer in the experimental setup. The PBS in the polarization analyzer helps distinguish the states $\ket{0}$ and $\ket{1}$. Her measurement results imply the following Bob's qubit states prepared by Alice via the RSP protocol in the ideal case of Eq.~(\ref{explain1}): $\ket{0}\rightarrow U\ket{1}$ and $\ket{1}\rightarrow U\ket{0}$.

In practical experimental settings with imperfect implementations, including the operation of  $U^\dag$, the sharded entangled pair, and the measurement in the experimental basis $\{\ket{i}\}$, the above RSP process is replaced by an experimental process $\mathcal{E}_{\text{expt}}$, which maps Alice's postmeasurement state to a state of Bob's qubit by
\begin{equation}
\rho_{\text{out},s^{\perp}}=\mathcal{E}_{\text{expt}}(\rho_{\text{in},s}) \text{  and  } \rho_{\text{out},s}=\mathcal{E}_{\text{expt}}(\rho_{\text{in},s^{\perp}}),
\end{equation} 
where $\rho_{\text{in},s}$ and $\rho_{\text{in},s^{\perp}}$ denote the experimental input states prepared by Alice's operations and measurements, while $\rho_{\text{out},s}$ and $\rho_{\text{out},s^{\perp}}$ are the resulting states that Bob can obtain. For the ideal case, $\rho_{\text{in},s}=\rho_{\text{out},s}=\ketbra{s}{s}$ and $\rho_{\text{in},s^{\perp}}=\rho_{\text{out},s^{\perp}}=\ketbra{s^{\perp}}{s^{\perp}}$.

\begin{figure}[t]
	\includegraphics[width=6cm]{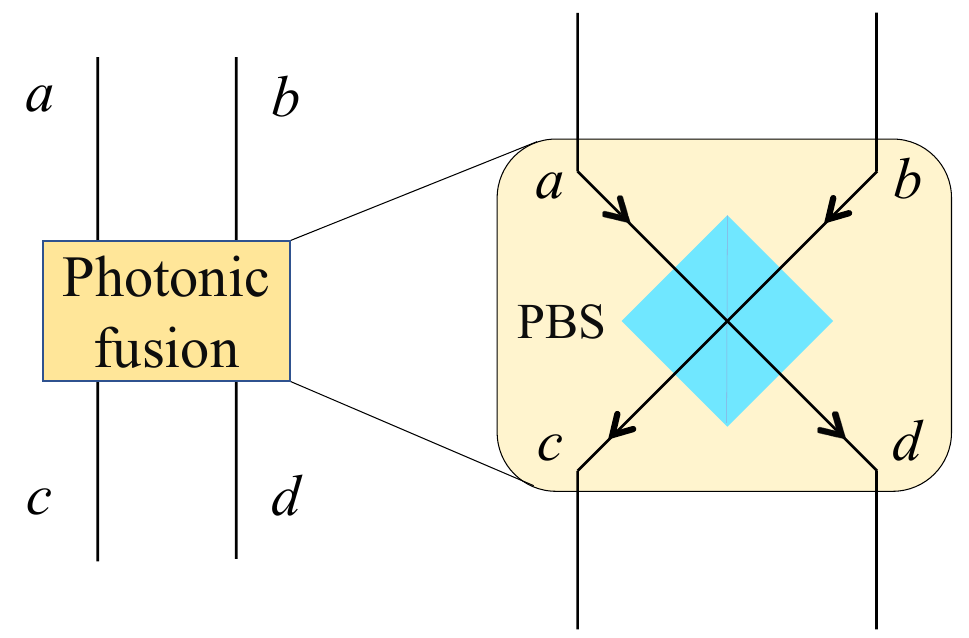}
	\caption{Photonic fusion unit. Two individual photons in the spatial modes $a$ and $b$ are input into a polarizing beam splitter (PBS) and post-selected in the spatial modes $c$ and $d$. To achieve temporal and spatial overlap at the PBS, the path of photon $b$ is adjusted by fine-tuning the delay with a nanopositioning stage in our experiment (not shown).}
	\label{fig:fusion}
\end{figure}

\subsection{Photonic fusion}

The fusion of photon pairs first coherently interferes with two individual photons in two different spatial modes ($a$ and $b$) at a PBS and then post-selects the two outputs in different modes ($c$ and $d$), as shown in the inset of Fig.~\ref{fig:fusion}. The PBS transmits horizontally polarized light ($H$) and reflects vertically polarized light ($V$). That is, after the PBS, the two-photon states $\ket{H_aH_b},$ $\ket{H_aV_b},$ $\ket{V_aH_b},$ $\ket{V_aV_b}$ become $\ket{H_dH_c},$ $\ket{H_dV_d},$ $\ket{V_cH_c},$ $\ket{V_cV_d}$, respectively, where ${H_i}$ (${V_i}$) denotes $H$ ($V$) polarization in the spatial mode $i=a,$ $b,$ $c,$ $d$. A subsequent post-selection in the modes, $c$ and $d$, picks the states $\ket{H_cH_d}$ and $\ket{V_cV_d}$ only, which makes the fusion process of the photon pairs non-trace preserving.

Since the photonic fusion superposes the two-photon amplitudes of different yet indistinguishable alternatives of the joint photodetection event in the different output modes of PBS, a quantum operation can represent an ideal photon fusion as described below~\cite{Lu07,Pan12,Wang16,Zhong18,Bell12}:
\begin{equation}\label{ideal_fusion}
\tilde{\rho}_\text{out}=\tilde{\mathcal{E}}_\text{fusion}(\rho_\text{in})=M\rho_\text{in}M^\dagger,
\end{equation}
where $\rho_\text{in}$ and $\tilde{\rho}_\text{out}$ denote the input and output states of the photon fusion $\tilde{\mathcal{E}}_\text{fusion}$, respectively, and
\begin{equation}\label{M}
M=\ket{H_{d}H_c}\!\!\bra{H_aH_b}+\ket{V_cV_d}\!\!\bra{V_aV_b},
\end{equation}
is called the fusion operator. Here $\tilde{\rho}_\text{out}$ is subnormalized, and its normalization factor ${\rm tr}(\tilde{\rho}_\text{out})$ is the probability of joint photodetection event. As will be shown below, we use $\tilde{\mathcal{E}}_\text{fusion}$ in the coherence transfer process. 

\subsection{Scheme for measurement-based one-qubit coherence transfer in multi-photon entanglement networks}

\begin{figure*}[t]
\includegraphics[width=13cm]{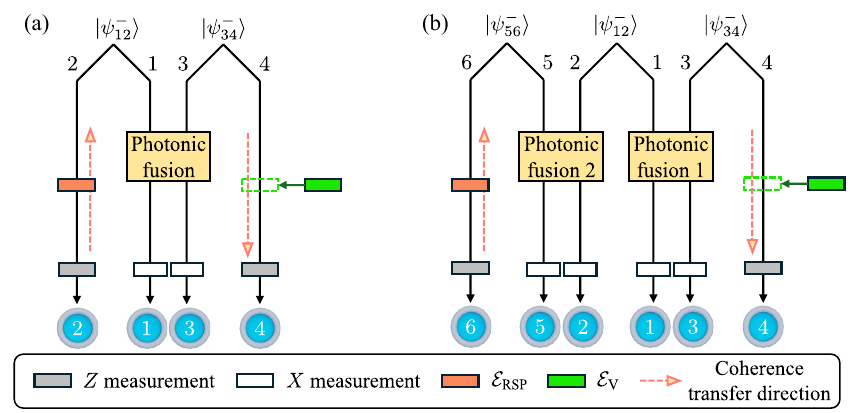}
\caption{One-qubit coherence transfer scheme and its detection method in (a) four-node and (b) six-node entanglement networks. We utilize RSP and photonic fusion to transfer the coherence of a single qubit through $N$-photon entanglement. For $N=4$, one RSP and one photonic fusion are used. For $N=6$, based on the case of $N=4$, an additional photonic fusion is used.}
\label{fig:n46}
\end{figure*}

We utilize RSP and photonic fusion to transfer the coherence of a single qubit through multi-photon entanglement. As shown in Fig.~\ref{fig:n46}(a), using RSP for the entangled state $\ket{\psi^{-}}_{12}$, assume photon 1 is prepared in $\rho_1$ of the form
\begin{eqnarray}
\begin{aligned}
{\rho_1}= 
\begin{bmatrix} 
\rho_{11} & \rho_{12}\\
\rho_{21} & \rho_{22}
\end{bmatrix}.
\end{aligned}
\end{eqnarray}
A subsequent photonic fusion of photon 1 and photon 3, followed by measurement in the $X$ basis, results in the state of photon 4 becoming the following density matrix:
\begin{eqnarray}
{\rho_4}&=&\text{tr}_{13}[\tilde{\mathcal{E}}_{\mathrm{fusion},13}(\rho_{1} \otimes \ket{\psi^{-}}_{3434}\!\bra{\psi^{-}})\ket{ij}_{1313\!}\bra{ij}]\nonumber\\
&=&\begin{bmatrix} 
\rho_{22} & (-1)^{k}\rho_{21}\\
(-1)^{k}\rho_{12} & \rho_{11}
\end{bmatrix},
\end{eqnarray}
for $i,j=+,-$, where $k=0$ for $i=j$, and $k=1$ otherwise. This result clearly indicates that a qubit's coherence remains unchanged after a photonic fusion operation. The states $\rho_1$ and $\rho_4$ are equivalent under a local operation.

\subsection{Experimental one-qubit coherence transfer and detection of coherence transfer in an $N$-node entanglement network}\label{sdp}

\begin{figure*}[t]
	\includegraphics[width=13cm]{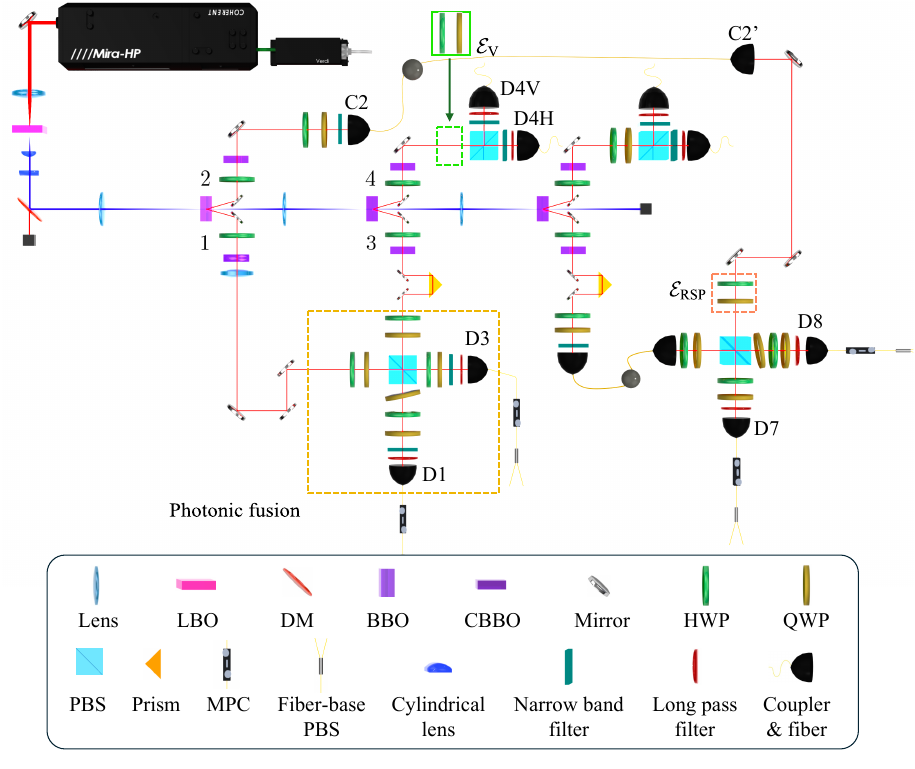}
	\caption{Experimental setup for one-qubit coherence transfer and detection in the four-photon quantum network. We generate three pairs of entangled photons. To transfer one-qubit coherence, we implement RSP using waveplates and polarization measurements in the $Z$ basis (laboratory basis) for input and transfer. To extend coherence transfer from two to four quantum nodes, the one photonic fusion operation is used. At the checkpoint node, the waveplate set and polarizer (or PBS) are combined to realize the operation $\mathcal{E}_{\text{V}}$.}
	\label{fig:OneQubitExptN4}
\end{figure*}

\begin{figure*}[t]
	\includegraphics[width=13cm]{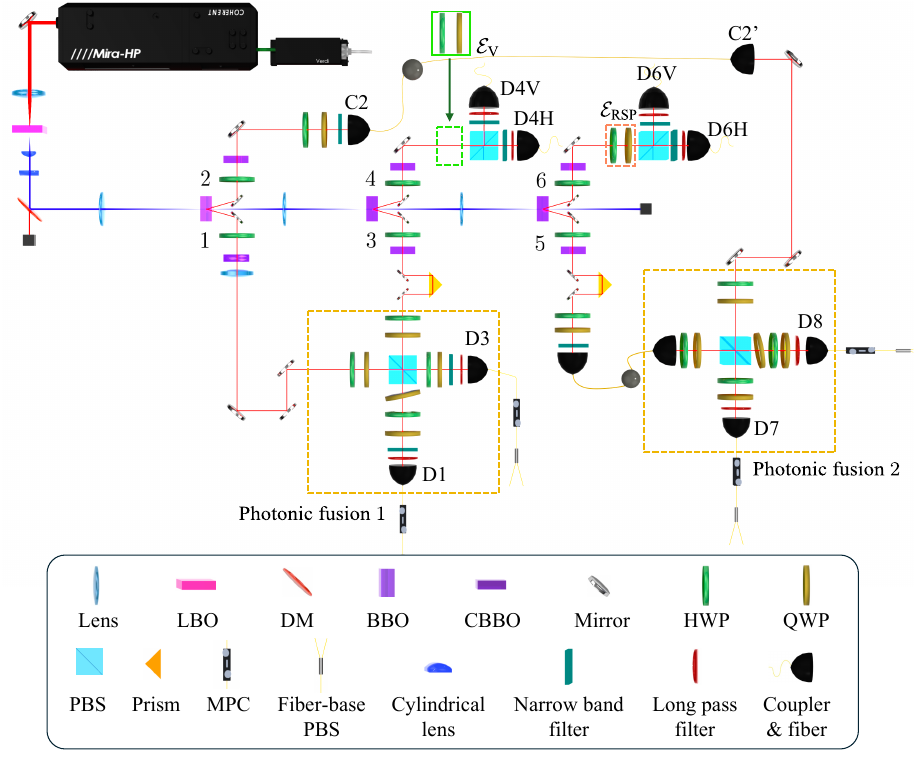}
	\caption{Experimental setup for one-qubit coherence transfer and detection in the six-photon quantum network. We generate three pairs of entangled photons. To transfer one-qubit coherence, we implement RSP using waveplates and polarization measurements in the $Z$ basis (laboratory basis) for input and transfer. To extend coherence transfer from two to six quantum nodes, we use two photonic fusion operations. At the checkpoint node, the waveplate set and polarizer (or PBS) are combined to realize the operation $\mathcal{E}_{\text{V}}$.}
	\label{fig:OneQubitExptN6}
\end{figure*}

\begin{figure}[t]
	\includegraphics[width=5.5cm]{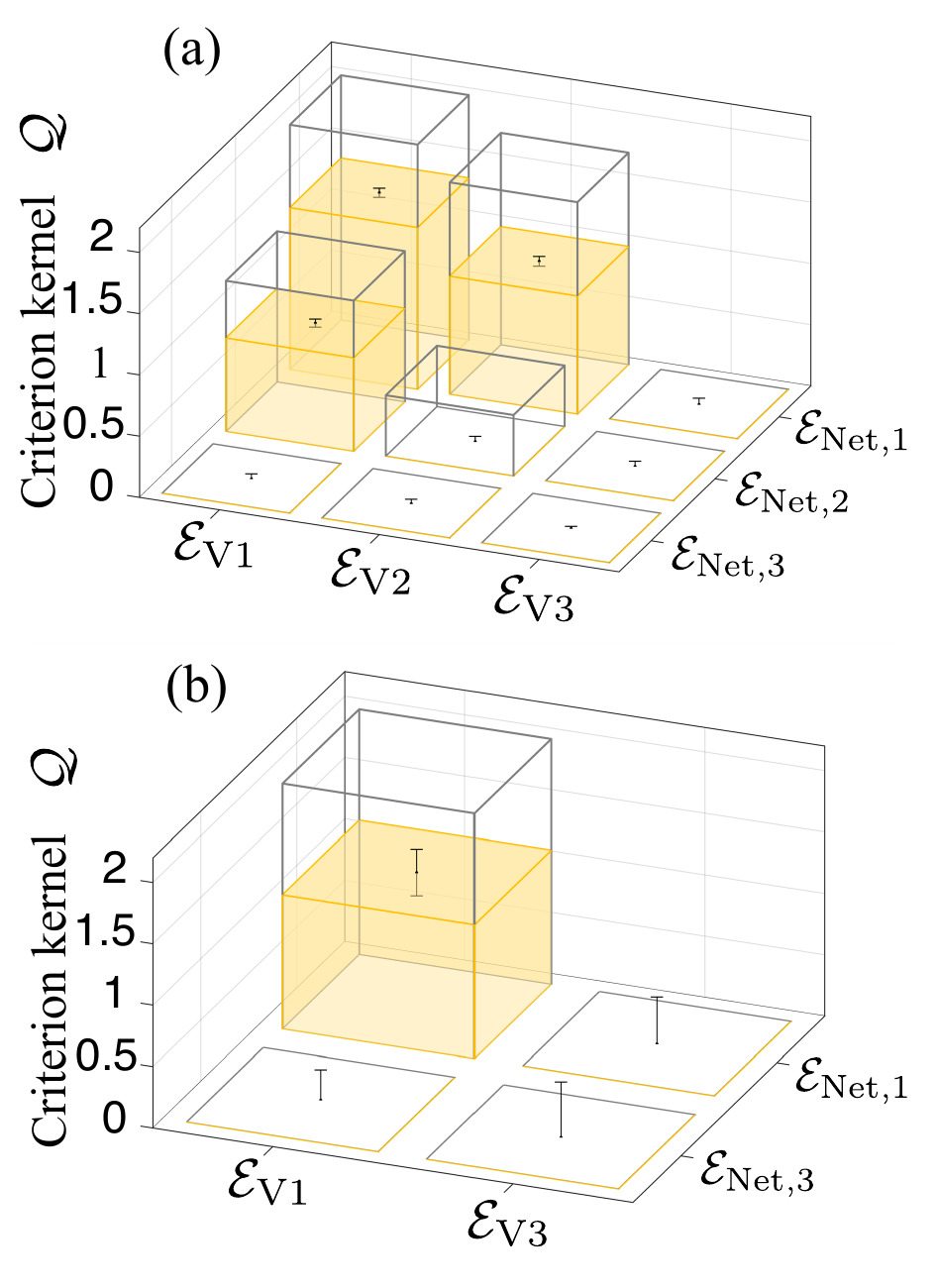}
	\caption{Experimental one-qubit coherence transfer. (a) Coherence transfer in a four-photon entanglement network and (b) coherence transfer in a six-photon entanglement network. We perform different coherence transfer processes, $\mathcal{E}_{\text{Net},j}$, which include the respective coherence inputs, $\mathcal{E}_{\text{RSP}j}$, for $j=1,2,3$, and the same photonic fusion processes, $\tilde{\mathcal{E}}_\text{fusion}$. The $\mathcal{E}_{\text{RSP}1}$, $\mathcal{E}_{\text{RSP}2}$, and $\mathcal{E}_{\text{RSP}3}$ correspond to the QWP set with the fast axis at $45^{\circ}$, $30^{\circ}$, and $0^{\circ}$ relative to the horizontal axis, respectively, while the HWP always sets the fast axis at $0^{\circ}$ relative to the horizontal axis. According to criterion~(1) in the main text, the operations $\mathcal{E}_{\text{V}j}$ applied to output states at the checkpoint node use the same HWP and QWP settings as in $\mathcal{E}_{\text{RSP}j}$. The ideal value of $\mathcal{Q}$ is shown by the gray solid line, while the experimentally measured value is depicted by the yellow bar. The main difference between the ideal and experimental values arises from limited state fidelity and errors in photonic fusion. Each experimental outcome is listed in Table~\ref{q_values_one_qubit}.}
	\label{onequbitresults}
\end{figure}

The experimental procedure for transferring and detecting coherence in an $N$-photon entanglement network is outlined as follows: input coherence, evolution, output, and detection.

1. Coherence input. We implement the RSP protocol to input a state $\rho_{\text{in},k}$, which is the post-measurement state of the input node, performed by the input node holder Alice, who implements the process $\mathcal{E}_{\text{RSP}}$ according to the target operation $U^{\dag}$ and the followed $Z$ measurement in the basis $\{\ket{i}\}=\{\ket{0}=\ket{H},\ket{1}=\ket{V}\}$. Photons 2 and 6 are measured to determine the input coherence of a single qubit at the input node for $N=4$ [Fig.~\ref{fig:n46}(a)] and $N=6$ [Fig.~\ref{fig:n46}(b)], respectively. For the ideal case, the input state are $\rho_{\text{in},k}=U\ket{k}\!\!\bra{k}U^{\dag}$ for $k=0,1$. Experimentally, we use three different settings of HWP and QWP to implement $U^{\dag}$, $\mathcal{E}_{\text{RSP}j}$, at the input node (Alice) for three experimental RSP processes. $\mathcal{E}_{\text{RSP}1}$, $\mathcal{E}_{\text{RSP}2}$, and $\mathcal{E}_{\text{RSP}3}$ correspond to QWP set with the fast axis at $45^{\circ}$, $30^{\circ}$, and $0^{\circ}$ relative to the horizontal axis, respectively, while HWP always sets the fast axis at $0^{\circ}$ relative to the horizontal axis. The $Z$ measurement at the input node is performed using a PBS. These individual coherence input processes, $\mathcal{E}_{\text{RSP}j}$, are used in each multi-photon entanglement network to transfer coherence, $\mathcal{E}_{\text{Net},j}$. The experimental setups for one-qubit coherence transfer and detection in the four-photon and six-photon quantum network are shown in Figs. \ref{fig:OneQubitExptN4} and \ref{fig:OneQubitExptN6}, respectively.

2. The evolution of coherence. To transmit the input state further in a four-photon entanglement network for $N=4$ [Fig.~\ref{fig:n46}(a)], we perform a photonic fusion $\tilde{\mathcal{E}}_\text{fusion}$ of photon 1, which is in the state after RSP at node 2, and photon 3 at node 3, along with the $X$ measurements in the basis $\{\ket{+},\ket{-}\}$ for both photons, considering the events where the post-measurement state is $\ket{+}$. This outputs a photon qubit of photon 4 at the output node. Similarly, for $N=6$ [Fig.~\ref{fig:n46}(b)], we perform a photonic fusion of photon 5, which is in the state after RSP at node 5, and photon 2 at node 2, along with the $X$ measurements for both photons, considering the events where the post-measurement state is $\ket{+}$. We perform a simultaneous and subsequent photonic fusion of photons 1 and 3 to transfer the qubit's coherence to photon 4 at output node 4. The photonic fusion is controlled by a prism with a nanopositioning linear stage, ensuring the interferometer is temporally synchronized. For a given $N$-photon network, the same photonic fusion processes $\tilde{\mathcal{E}}_\text{fusion}$ are used in all three coherence transfer processes $\mathcal{E}_{\text{Net},j}$, each including the respective coherence input $\mathcal{E}_{\text{RSP}j}$, for $j=1,2,3$.

3. Output of coherence. We measure photon 4 at the output checkpoint nodes 4 and 6 in the output states of the four-photon [Fig.~\ref{fig:n46}(a)] and six-photon [Fig.~\ref{fig:n46}(b)] entanglement networks, $\rho_{\text{out},k}$, respectively, according to the introduced criterion. We perform the population measurement ($Z$ measurement) on the state $\rho_{\text{out},k}$ to obtain the diagonal elements of the density operator $\rho_{\text{out},k}^{(\text{d})}$. In each experimental coherence process $\mathcal{E}_{\text{Net},j}$, the operation at the checkpoint $\mathcal{E}_\text{V}=\mathcal{E}_{\text{V}j}$ is applied to $\rho_{\text{out},k}$ followed by a $Z$ measurement to obtain the probabilities $P_{\mathcal{E}_\text{V}}(i|\rho_{\text{out},k})$. In the experiment, the operations $\mathcal{E}_{\text{V}j}$ applied to output states at the checkpoint are performed using the same settings of HWP and QWP in $\mathcal{E}_{\text{RSP}j}$ for $j=1,2,3$ to demonstrate how $\mathcal{E}_V$'s coherence creation capability affects the detection of coherence transfer.

4. Detection of coherence transfer. With regarding the experimental probabilities $\rho_{\text{out},k}^{(\text{d})}$ and $P_{\mathcal{E}_\text{V}}(i|\rho_{\text{out},k})$ for both experimental coherence transfer processes in the cases of $N=4$ and $N=6$, we analyze the experimental transfer process using the introduced criterion, Eq.~(1) in the main text. To conduct a comprehensive transfer assessment, we employ the SDP technique~\cite{Lofberg2004Yalmip,Toh1999Sdpt3,Andersen2000Mosek} to find the minimum value of the criterion kernel. SDP identifies a process that satisfies all the provided constraints and determines the minimum value from an infinite set of possible options. The constraints we impose for calculating our criterion ensure that the process found by SDP complies with the process definition and cannot create coherence. To thoroughly construct the process matrix that describes an incapable process of coherence creation, $\mathcal{E}_{\mathcal{I}}$, we start by preparing the following input states for $\mathcal{E}_{\mathcal{I}}$ \cite{Kuo19}: 
\begin{equation}
	\ket{\psi_{kmn}} = 
	\begin{cases}
		\frac{1}{2} \big( \ket{m} + \ket{n} \big), & \text{if } k = 1,\, m = n, \\
		\frac{1}{\sqrt{2}} \big( \ket{m} + \ket{n} \big), & \text{if } k = 2,\, m \neq n, \\
		\frac{1}{\sqrt{2}} \big( \ket{m} + i\ket{n} \big), & \text{if } k = 3,\, m \neq n, 
	\end{cases}
\end{equation}
where $m, n = 0, 1, \dots, d-1$. The corresponding density matrix of the input state is given by $\rho_{\text{in}} = \rho_{kmn} = \ket{\psi_{kmn}}\bra{\psi_{kmn}}.$ From these, we obtain the output states $\mathcal{E}_\mathcal{I}(\rho_{kmn})$, which correspond to the prepared inputs. An arbitrary output state $\mathcal{E}_\mathcal{I}(\rho)$ can then be expressed as a linear combination of the output states $\mathcal{E}_\mathcal{I}(\rho_{kmn})$. The constraint set, representing the incapable process of creating coherence, is defined as follows:
\begin{equation}
	\begin{aligned}
		\mathcal{E}_\mathcal{I}(\rho_{kmn}) &\geq 0, 
		&& \forall\, k, m, n; \\[6pt]
		\mathcal{E}_\mathcal{I}(\rho_{1mm}) 
		&= \sum_{i=0}^{d-1} p_{m,i} \ketbra{i}{i} \in \widetilde{\mathcal{S}}_{\text{incoh}}, && \forall m.
	\end{aligned}
\end{equation}
where $p_{m,i} \geq 0$ are the non-negative coefficients of each pure state $\ketbra{i}{i}$, and $\widetilde{\mathcal{S}}_{\text{incoh}}$ denotes the set of unnormalized incoherent states with respect to the computational basis $\{\ket{i}\}$. These conditions guarantee that $\mathcal{E}_\mathcal{I}$ is both reasonable and incapable of creating coherence.

For $d=2$ as considered in the experimental detection of one-qubit coherence transfer, we need to consider the following input states:  $\ket{\psi_{kmn}}$ for $k=1,2,3$ and $m,n=0,1$, to define the incapable process $\mathcal{E}_{\mathcal{I}}$. Explicitly, that is: $\ket{\psi_{100}}=\ket{0}$, $\ket{\psi_{111}}=\ket{1}$, $\ket{\psi_{201}}=\ket{+}$, and $\ket{\psi_{301}}=\ket{R}$, where $\ket{+}=1/\sqrt{2}(\ket{0}+\ket{1})$ and $\ket{R}=1/\sqrt{2}(\ket{0}+i\ket{1})$. Since the corresponding output states of these four input states are sufficient to construct and define a process matrix, $\mathcal{E}_{\mathcal{I}}$ for $d=2$ is fully described by the following output states that $\mathcal{E}_{\mathcal{I}}$ should produce:
\begin{eqnarray}
&&\mathcal{E}_\mathcal{I}(\rho_{201})\geq 0, \mathcal{E}_\mathcal{I}(\rho_{301})\geq 0\nonumber\\
&&\mathcal{E}_\mathcal{I}(\rho_{1mm}) = \sum_{i=0}^{1} p_{m,i} \ketbra{i}{i} \in \widetilde{\mathcal{S}}_{\text{incoh}},\text{ for } m=0,1.\nonumber
\end{eqnarray}
The reasons the above output state descriptions define $\mathcal{E}_{\mathcal{I}}$ are that, for a process incapable of creating coherence, $\mathcal{E}_\mathcal{I}(\rho_{1mm})$ should remain incoherent, since the input states are incoherent. Moreover, in addition to positivity, there are no further constraints on the output states $\mathcal{E}_\mathcal{I}(\rho_{201})$ and $\mathcal{E}_\mathcal{I}(\rho_{301})$ when they are coherent before the process.

Subsequently, by taking the quantum coherence criterion in Eq.~(1) as the objective function and executing the corresponding optimization procedure \cite{Lofberg2004Yalmip,Toh1999Sdpt3,Andersen2000Mosek}, we can determine the optimal solution that minimizes the objective. The experimental detections of coherence transfer are shown in Fig.~\ref{onequbitresults}, with details of each detection result provided in Table~\ref{q_values_one_qubit}.

\section{Detection of two-qubit coherence transfer in multi-photon entanglement networks}

We use Bell state measurement (BSM) and entangled photon pairs to execute the three stages of measurement-based coherence transfer of two qubits: input coherence, evolution, and output, in our four-photon and six-photon entanglement quantum networks, as shown in Fig.~4 in the main text. Their fundamental principles are outlined below. Using these elements, we will explain how to transfer and detect two-qubit coherence in multi-photon entanglement networks. The two-qubit coherence transfer demonstrated here is comparable to entanglement swapping.

\begin{table*}[t]
\centering
\caption{Measured one-qubit coherence transfer kernel values of $\mathcal{Q}$, displayed in Figs.~\ref{onequbitresults}(a) four-photon network and (b) six-photon network.}
\label{q_values_one_qubit}
\vspace{0.5em}
\begin{tabular}{lcccc}
\toprule
\textbf{Network} & \textbf{Operation} &  $\mathcal{E}_{\text{Net,1}}$ &  $\mathcal{E}_{\text{Net,2}}$ &  $\mathcal{E}_{\text{Net,3}}$ \\
\midrule
\textbf{4 Photons } & $\mathcal{E}_{\text{V}1}$ & $1.3232 \pm 0.0374$ & $0.7652 \pm 0.0325$ & $0.0000 \pm 0.0408$ \\
 & $\mathcal{E}_{\text{V}2}$  & $0.9664 \pm 0.0400$ & $0.0000 \pm 0.0429$ & $0.0000 \pm 0.0355$\\
 & $\mathcal{E}_{\text{V}3}$  & $0.0000 \pm 0.0510$ & $0.0000 \pm 0.0426$ & $0.0000 \pm 0.0162$\\
\midrule
\textbf{6 Photons} & $\mathcal{E}_{\text{V}1}$ & $1.0944 \pm 0.1896$ &   & $0.0000 \pm 0.2458$ \\
 & $\mathcal{E}_{\text{V}3}$  & $0.0000 \pm 0.3825$ &   & $0.0000 \pm 0.4506$ \\
\bottomrule
\end{tabular}
\end{table*}

\subsection{Bell state measurement}

\subsubsection{Onto the states: $\{\ket{\phi^+},\ket{\phi^-}\}$}

A photonic fusion operation can effectively project onto the Bell basis \cite{Pan01,Browne05,Bodiya06}. However, in practical applications, implementing BSM with linear optics faces fundamental limitations: out of the four Bell states, only two can be reliably distinguished in a single setup using a PBS and photon detectors. The PBS operator for photonic fusion introduced earlier can be expressed as follows:
\begin{eqnarray}
M_{\text{PBS}} &=& \ket{H_{\mathrm{c}} H_{\mathrm{d}}}\!\! \bra{H_{\mathrm{a}} H_{\mathrm{b}}} + \ket{V_{\mathrm{c}} V_{\mathrm{d}}}\!\! \bra{V_{\mathrm{a}} V_{\mathrm{b}}} \nonumber\\
&=& \ket{\phi^+_{\mathrm{cd}}}\!\! \bra{\phi^+_{\mathrm{ab}}} + \ket{\phi^-_{\mathrm{cd}}}\!\! \bra{\phi^-_{\mathrm{ab}}},
\end{eqnarray}
where modes labeled a and b are the input ports, and modes c and d are the output ports of the PBS. It is important to note that this operator employs the post-selection method, which filters outcomes by requiring simultaneous detection across different modes. In practice, input states such as $\ket{HV}$ or $\ket{VH}$ are routed into the same output port, making them indistinguishable by coincidence detection and limiting the ability to identify all Bell states unambiguously.

\begin{figure}[t]
	\includegraphics[width=8.5cm]{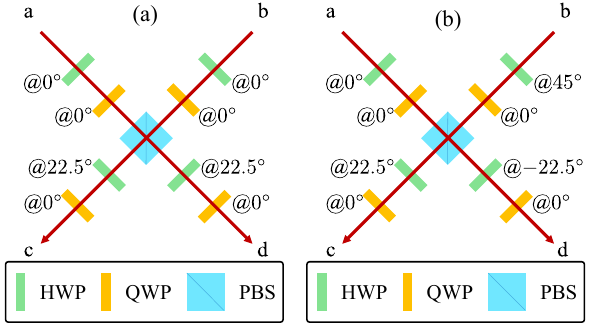}
	\caption{PBS-based BSM. (a) The first PBS configuration enables the BSM to project onto the Bell states $\ket{\phi^+}$ or $\ket{\phi^-}$. The waveplates configuration at the output modes c and d, together with the subsequent fiber-based PBS (not shown in the figure), enables measurement in the $X$ basis. (b) The second PBS configuration enables the BSM to project onto the Bell states $\ket{\psi^+}$ or $\ket{\psi^-}$. The waveplate configuration applied to modes a and b implements the local operation $I_{\mathrm{a}} \otimes \sigma_X^{\mathrm{b}}$. Specifically, the corresponding settings consist of $\{\text{HWP@0°},\, \text{QWP@0°}\}$ for mode a and $\{\text{HWP@(0°+45°)},\, \text{QWP@0°}\}$ for mode b. Likewise, the configuration for modes c and d, along with the subsequent fiber-based PBS (not shown in the figure), realizes the transformation $I_{\mathrm{c}} \otimes \sigma_X^{\mathrm{d}}$ and implements measurement in the $X$ basis. The associated waveplate settings are $\{\text{HWP@22.5°},\, \text{QWP@0°}\}$ for mode c and $\{\text{HWP@(22.5°–45°)},\, \text{QWP@0°}\}$ for mode d.}
	\label{fig:BSM}
\end{figure}

When this operator is applied to the input state $\ket{\psi_{\mathrm{ab}}}$, it produces the corresponding output state $\ket{\psi_{\mathrm{cd}}}$, up to a normalization constant, as shown below:
\begin{eqnarray}
\ket{\psi_{\mathrm{cd}}} 
	&=& M_{\text{PBS}} \ket{\psi_{\mathrm{ab}}}\nonumber\\
	&=& \braket{\phi^+_{\mathrm{ab}} | \psi_{\mathrm{ab}}} \ket{\phi^+_{\mathrm{cd}}} 
	+ \braket{\phi^-_{\mathrm{ab}} | \psi_{\mathrm{ab}}} \ket{\phi^-_{\mathrm{cd}}},
\end{eqnarray}
The Bell states $\ket{\phi^+}$ and $ \ket{\phi^-}$ can be expressed in the $X$ basis $(\ket{+},\ket{-})$ as follows:
\begin{eqnarray}
\ket{\phi^+} &=& \frac{1}{\sqrt{2}} \left( \ket{++} + \ket{--} \right),\nonumber\\
\ket{\phi^-} &=& \frac{1}{\sqrt{2}} \left( \ket{+-} + \ket{-+} \right).\nonumber
\end{eqnarray}
Substituting these expressions into the state $\ket{\psi_{\mathrm{cd}}} $, we obtain the following result:
\begin{equation}
\begin{aligned}
		\ket{\tilde{\psi}_{\mathrm{cd}}}
		&= \braket{\phi^+_{\mathrm{ab}} | \psi_{\mathrm{ab}}} \cdot \frac{1}{\sqrt{2}} \left( \ket{+_{\mathrm{c}} +_{\mathrm{d}}} + \ket{-_{\mathrm{c}} -_{\mathrm{d}}} \right) \\
		&\quad + \braket{\phi^-_{\mathrm{ab}} | \psi_{\mathrm{ab}}} \cdot \frac{1}{\sqrt{2}} \left( \ket{+_{\mathrm{c}} -_{\mathrm{d}}} + \ket{-_{\mathrm{c}} +_{\mathrm{d}}} \right).
	\end{aligned}
\end{equation}
From the expression of $\ket{\tilde{\psi}_{\mathrm{cd}}}$ in the diagonal basis, it follows that a measurement outcome of $\ket{+_{\mathrm{c}} +_{\mathrm{d}}}$ or $\ket{-_{\mathrm{c}} -_{\mathrm{d}}}$ corresponds to a projection onto the Bell state $\ket{\phi^+_{\mathrm{ab}}}$. Similarly, outcomes $\ket{+_{\mathrm{c}} -_{\mathrm{d}}}$ or $\ket{-_{\mathrm{c}} +_{\mathrm{d}}}$ correspond to a projection onto $\ket{\phi^-_{\mathrm{ab}}}$. Therefore, by performing measurements in the diagonal polarization basis on the output modes c and d following the PBS, it is possible to project onto two of the four Bell states, $\ket{\phi^+_{\mathrm{ab}}}$ and $\ket{\phi^-_{\mathrm{ab}}}$, as shown in Fig.~\ref{fig:BSM}(a).

\subsubsection{Onto the states: $\{\ket{\psi^+},\ket{\psi^-}\}$}

\begin{figure*}[t]
	\includegraphics[width=12cm]{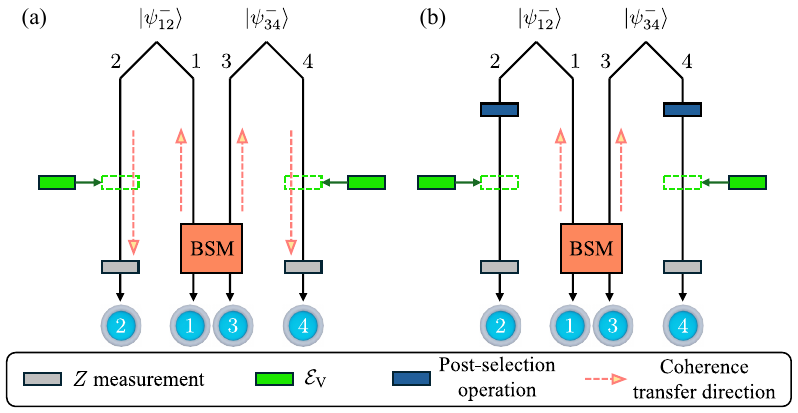}
	\caption{Two-qubit coherence transfer scheme and its detection method in four-node entanglement networks. (a) A process that transfers coherence within the quantum network is achieved by inputting coherence, implemented through a BSM. To measure the criterion kernel, $Z$ basis measurements are performed either directly on the output of $\rho_{\text{out},k}$, resulting in the state $\rho_{\text{out},k}^{(\text{d})}$, or after applying an additional experimental checkpoint operation $\mathcal{E}_{\text{V}}$, which is used to determine the conditional probability $P_{\mathcal{E}_\text{V}}(i|\rho_{\text{out},k})$. (b) As a control experiment, a process following the coherence input from the BSM suppresses coherence transfer, modeling a process incapable of such transfer in a network.}
	\label{fig:TwoQubitN4}
\end{figure*}

We now examine a different configuration of the PBS that allows the BSM to project onto the Bell states $\ket{\psi^+}$ and $ \ket{\psi^-}$. Applying the identity operator $I$ and the Pauli $X$ operator $\sigma_X$ to specific modes corresponds to local unitary transformations that rotate the polarization basis, which can be physically implemented using appropriate arrangements of HWP and QWP. Specifically, we consider the transformation where $I$ is applied to input mode a and $\sigma_X$ to input mode b, and similarly, $I$ is applied to output mode c and $\sigma_X$ to output mode d. Under these local operations, the PBS operator transforms into the following form.
\begin{equation}
M_{\text{PBS}}' = (I_{\mathrm{c}} \otimes \sigma_X^{\mathrm{d}}) \, M_{\text{PBS}} \, (I_{\mathrm{a}} \otimes \sigma_X^{\mathrm{b}}).
\end{equation}
Replacing the original PBS operator with the expression above yields the result:
\begin{equation}
\begin{aligned}
		M_{\text{PBS}}' 
		&= (I_{\mathrm{c}} \otimes \sigma_X^{\mathrm{d}}) \left( \ket{\phi^+_{\mathrm{cd}}} \bra{\phi^+_{\mathrm{ab}}} + \ket{\phi^-_{\mathrm{cd}}} \bra{\phi^-_{\mathrm{ab}}} \right) (I_{\mathrm{a}} \otimes \sigma_X^{\mathrm{b}}) \\
		&= \ket{\psi^+_{\mathrm{cd}}} \bra{\psi^+_{\mathrm{ab}}} + \ket{\psi^-_{\mathrm{cd}}} \bra{\psi^-_{\mathrm{ab}}}.
	\end{aligned}
\end{equation}
Likewise, the output state resulting from applying this transformation to the input state $\ket{\psi_{\text{ab}}}$, up to a normalization constant, is given by the state
\begin{eqnarray}
\ket{\tilde{\psi}_{\mathrm{cd}}} 
	&=& M_{\text{PBS}}' \ket{\psi_{\mathrm{ab}}}\\
	&=& \braket{\psi^+_{\mathrm{ab}} | \psi_{\mathrm{ab}}} \ket{\psi^+_{\mathrm{cd}}} 
	+ \braket{\psi^-_{\mathrm{ab}} | \psi_{\mathrm{ab}}} \ket{\psi^-_{\mathrm{cd}}},
\end{eqnarray}
The Bell states $\ket{\psi^+}$ and $\ket{\psi^-}$ can be expressed in this form.
\begin{eqnarray}
\ket{\psi^+} &=& \frac{1}{\sqrt{2}} \left( \ket{++} - \ket{--} \right),\nonumber\\
\ket{\psi^-} &=& \frac{1}{\sqrt{2}} \left( \ket{+-} - \ket{-+} \right).\nonumber
\end{eqnarray}
Then, we obtain the following result:
\begin{equation}
\begin{aligned}
		\ket{\tilde{\psi}_{\mathrm{cd}}}
		&= \braket{\psi^+_{\mathrm{ab}} | \psi_{\mathrm{ab}}} \cdot \frac{1}{\sqrt{2}} \left( \ket{+_{\mathrm{c}} +_{\mathrm{d}}} - \ket{-_{\mathrm{c}} -_{\mathrm{d}}} \right) \\
		&\quad + \braket{\psi^-_{\mathrm{ab}} | \psi_{\mathrm{ab}}} \cdot \frac{1}{\sqrt{2}} \left( \ket{+_{\mathrm{c}} -_{\mathrm{d}}} - \ket{-_{\mathrm{c}} +_{\mathrm{d}}} \right).
	\end{aligned}
\end{equation}
From the expression of $\ket{\tilde{\psi}_{\mathrm{cd}}}$ in the diagonal basis, it follows that a measurement outcome of $\ket{+_{\mathrm{c}} +_{\mathrm{d}}}$ or $\ket{-_{\mathrm{c}} -_{\mathrm{d}}}$ corresponds to a projection onto the Bell state $\ket{\psi^+_{\mathrm{ab}}}$. Similarly, outcomes $\ket{+_{\mathrm{c}} -_{\mathrm{d}}}$ or $\ket{-_{\mathrm{c}} +_{\mathrm{d}}}$ correspond to a projection onto $\ket{\psi^-_{\mathrm{ab}}}$. Consequently, by appropriately applying local unitary operations before and after the PBS, implemented using combinations of HWPs and QWPs, the set of distinguishable Bell states can be effectively expanded. This allows the BSM to project onto two of the four Bell states, $\ket{\psi^+_{\mathrm{ab}}}$ and $\ket{\psi^-_{\mathrm{ab}}}$, as shown in Fig.~\ref{fig:BSM}(b).

Although BSMs implemented with linear optics are inherently limited, specific configurations using PBS, along with local unitary transformations, can probabilistically distinguish all four Bell states. This ability is essential for executing network operations as outlined in our coherence transfer criterion framework. In the next subsection, we will demonstrate how BSM enables the input of two-qubit coherence across the network.

\begin{figure*}[t]
	\includegraphics[width=13cm]{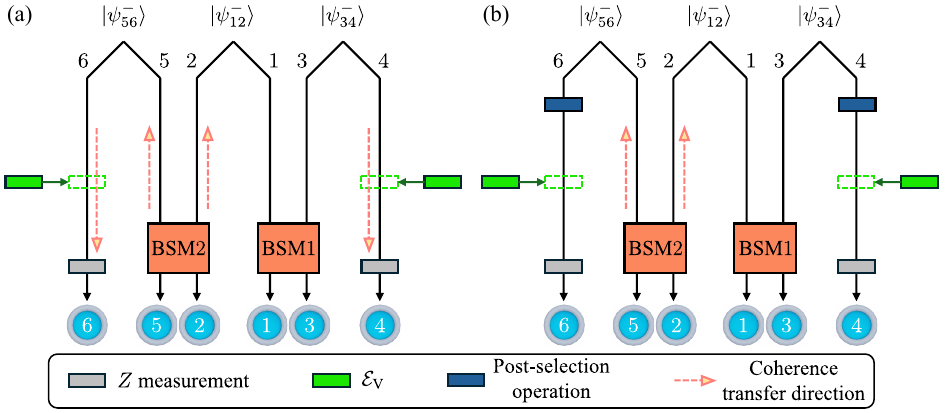}
	\caption{Two-qubit coherence transfer scheme and its detection method in six-node entanglement networks. (a) A process that transfers coherence within the quantum network is achieved by inputting coherence via a BSM and a second BSM for transmitting it to the two checkpoint nodes. To measure the criterion kernel, $Z$ basis measurements are performed either directly on the output of $\rho_{\text{out},k}$, resulting in the state $\rho_{\text{out},k}^{(\text{d})}$, or after applying an additional experimental checkpoint operation $\mathcal{E}_\text{V}$, which is used to determine the conditional probability $P_{\mathcal{E}_\text{V}}(i|\rho_{\text{out},k})$. (b) As a control experiment, a selection process following the BSM's coherence input suppresses coherence transfer, modeling a process incapable of such transfer in a network.}
	\label{fig:TwoQubitN6}
\end{figure*}

\subsection{Scheme for measurement-based two-qubit coherence transfer in multi-photon entanglement networks}

We use BSM and entangled photon pairs to transfer the coherence of two qubits in multi-photon entanglement networks. The result of the BSM on qubits 1 and 3 clearly determines the entangled state of qubits 2 and 4. From a coherence-transfer perspective, the coherence of the Bell states is input via the BSM. Meanwhile, the entanglement between nodes ensures the transfer of coherence to a subsequent pair, even without direct interaction between them. This is an entanglement-swapping \cite{Zukowski93,Pan98ES,Pan98} process that confirms that entanglement has been successfully swapped between qubits 2 and 4. See Fig.~\ref{fig:TwoQubitN4}(a).

To extend the coherence transfer from four-photon to six-photon networks, we add an extra BSM process and one more entangled photon pair to further transfer the two-qubit coherence. This additional BSM serves a role similar to its role in one-qubit coherence transfer. Both the BSM and photonic fusion operations rely on the interference of two indistinguishable photons and on post-selection via photon measurement. See Fig.~\ref{fig:TwoQubitN6}(a).

\subsection{Experimental two-qubit coherence transfer and detection of coherence transfer in an $N$-node entanglement network}

\begin{figure*}[t]
	\includegraphics[width=13cm]{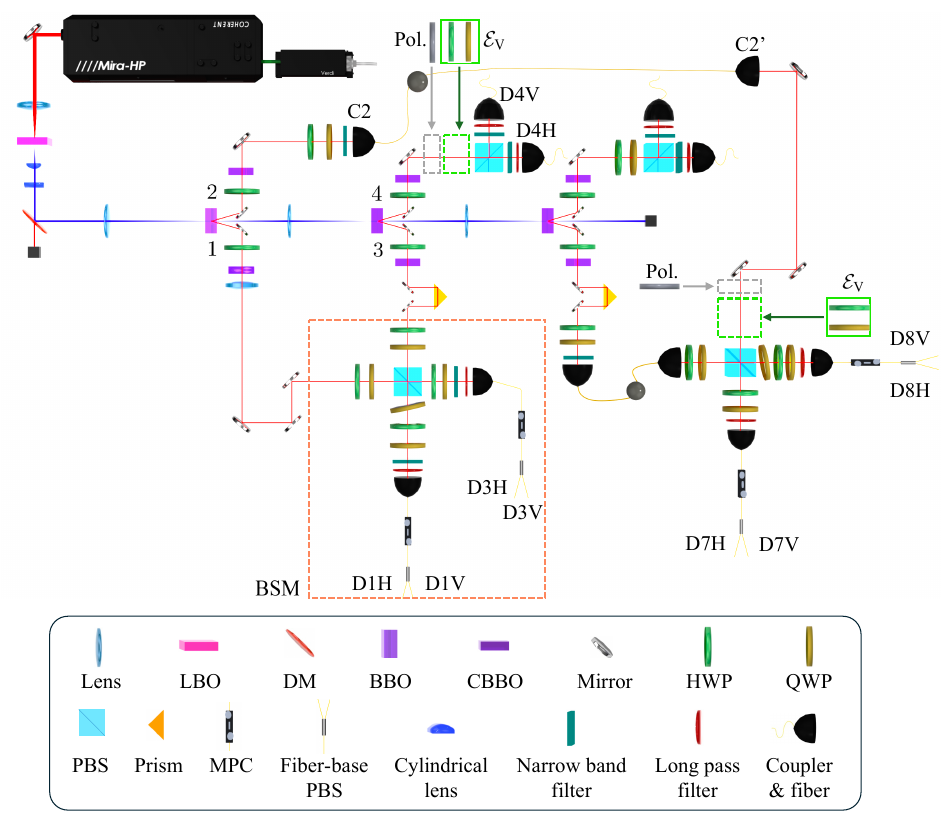}
	\caption{Experimental setup for two-qubit coherence transfer and detection in the four-photon quantum network. A process capable of transferring coherence within the quantum network is achieved through BSM and shared entanglement. Two entangled photon pairs are produced, and a BSM on photons 1 and 3 allows for coherence input and transfer to photons 2 and 4 ($\mathcal{E}_{\text{Net},1}$). Coherence creation is indicated by three operations, $\mathcal{E}_{\text{V}1}$, $\mathcal{E}_{\text{V}2}$, and $\mathcal{E}_{\text{V}3}$, using waveplates with QWP angles at 45°, 30°, and 0°. A control setup ($\mathcal{E}_{\text{Net},2}$) suppresses coherence via polarizers
	(Pol.); see Fig.~\ref{fig:TwoQubitN4}(b). The value of the criterion kernel $\mathcal{Q}$ is determined via SDP. The experimental detection results are presented in Fig.~3(d) in the main text.}
	\label{fig:TwoQubitExptN4}
\end{figure*}

\begin{figure*}[t]
	\includegraphics[width=13cm]{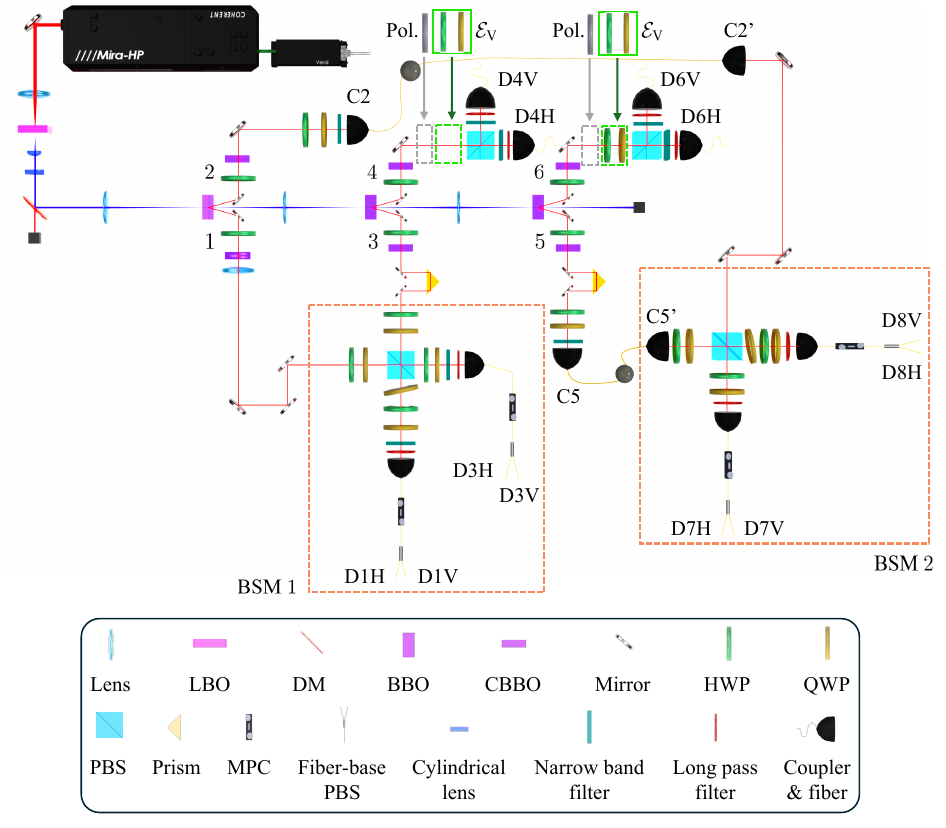}
	\caption{Experimental setup for two-qubit coherence transfer and detection in the six-photon quantum network. A process that transfers coherence within the quantum network is achieved through BSM and shared entanglement. Three entangled photon pairs are generated, and BSMs on qubit pairs (5, 2) and (1, 3) enable inputting and transferring coherence, respectively, resulting in coherence transfer from photons 1 and 3 to photons 6 and 4, $\mathcal{E}_{\text{Net},1}$. Coherence transfer processes are detected by three operations, $\mathcal{E}_{\text{V}1}$, $\mathcal{E}_{\text{V}2}$, and $\mathcal{E}_{\text{V}3}$, using wave plates with QWP angles at 45°, 30°, and 0°. A control setup resulting in a process, $\mathcal{E}_{\text{Net},2}$, suppresses coherence using polarizers (Pol.); see Fig.~\ref{fig:TwoQubitN6}(b). The value of the criterion kernel $\mathcal{Q}$ is determined via SDP. The experimental detection results are presented in Fig.~3(e) in the main text.}
	\label{fig:TwoQubitExptN6}
\end{figure*}

The experimental procedure for transferring and detecting coherence in an $N$-photon entanglement network is outlined as follows: input coherence, evolution, output, and detection.

1. Input and the evolution of coherence. We implement the BSM protocol to input a state $\rho_{\text{in},k}$ with Bell-state coherence, which is the post-measurement state of the input node, measured by two input node holders. BSMs are carried out on photons 1 and 3 [Fig.~\ref{fig:TwoQubitN4}(a)]. We aim to input the following states in the ordering:
\begin{eqnarray}
		\rho_{\text{in},1} &= \ketbra{\psi^+}{\psi^+}, \nonumber\\
		\rho_{\text{in},2} &= \ketbra{\psi^-}{\psi^-}, \\
		\rho_{\text{in},3} &= \ketbra{\phi^+}{\phi^+}, \nonumber\\
		\rho_{\text{in},4} &= \ketbra{\phi^-}{\phi^-}.\nonumber
\end{eqnarray}
As previously detailed, the first PBS setup enables projection onto the Bell states $\ket{\phi^+}$ or $\ket{\phi^-}$, while the second setup allows projection onto $\ket{\psi^+}$ or $\ket{\psi^-}$. For four-photon quantum networks, this process transfers the coherence of each of the four corresponding Bell states to photons 2 and 4 [Fig.~\ref{fig:TwoQubitN4}(a)]. In six-photon quantum networks, building on the four-photon entanglement network, an additional BSM transfers the two-qubit coherence from the third and fourth nodes to the fifth (mode 4) and sixth (mode 6) nodes [Fig.~\ref{fig:TwoQubitN6}(a)]. For both $N=4$ [Fig.~\ref{fig:TwoQubitN4}(a)] and $N=6$ [Fig.~\ref{fig:TwoQubitN6}(a)], these processes are considered capable of coherence transfer and are denoted by $\mathcal{E}_{\text{Net},1}$. The experimental setup for two-qubit coherence transfer and detection in the four-photon and six-photon quantum network is shown in Figs. \ref{fig:TwoQubitExptN4} and \ref{fig:TwoQubitExptN6}, respectively.

2. Output of coherence. In an ideal scenario, we anticipate observing the following output states.
\begin{eqnarray}
		\rho_{\text{out},1} &= \ketbra{\psi^+}{\psi^+}, \nonumber\\
		\rho_{\text{out},2} &= \ketbra{\psi^-}{\psi^-}, \\
		\rho_{\text{out},3} &= \ketbra{\phi^+}{\phi^+}, \nonumber\\
		\rho_{\text{out},4} &= \ketbra{\phi^-}{\phi^-}.\nonumber
\end{eqnarray}
We measure photons (2,4) at the output checkpoint nodes (2,4), and photons (6,4) at the output checkpoint nodes (6,4) in the output states of the four-photon [Fig.~\ref{fig:TwoQubitN4}(a)] and six-photon [Fig.~\ref{fig:TwoQubitN6}(a)] entanglement networks, respectively, according to the introduced criterion~(1) in the main text. We perform the population measurement ($Z$ measurement) on the two-photon state $\rho_{\text{out},k}$ to obtain the diagonal elements of the density operator $\rho_{\text{out},k}^{(\text{d})}$. In each experimental coherence process $\mathcal{E}_{\text{Net},1}$, the operation at the checkpoint $\mathcal{E}_{\text{V}}=\mathcal{E}_{\text{V}j}$ is applied to $\rho_{\text{out},k}$ followed by a $Z$ measurement to obtain the probabilities $P_{\mathcal{E}_\text{V}}(i|\rho_{\text{out},k})$ for $i=0,1,2,3$. Here, we define the horizontal and vertical polarization states of a photon that form the laboratory basis, $\{\ket{i}\}=\{\ket{0}=\ket{HH},\ket{1}=\ket{HV},\ket{2}=\ket{VH},\ket{3}=\ket{VV}\}$. In the experiment, the operations $\mathcal{E}_{\text{V}}=\mathcal{E}_{\text{V}j}$ applied to output states at the checkpoint are performed using the same settings of HWP and QWP in $\mathcal{E}_{\text{RSP}j}$ for $j=1,2,3$ as used in detecting one-qubit coherence transfer, to demonstrate how $\mathcal{E}_{\text{V}}$'s coherence creation capability affects the detection of coherence transfer. For each identification process, a set of waveplates is used, including a QWP with its fast axis oriented at 45° ($\mathcal{E}_{\text{V}1}$), 30° ($\mathcal{E}_{\text{V}2}$), and 0° ($\mathcal{E}_{\text{V}3}$) relative to the horizontal axis. The HWP is always set with its fast axis at 0°. The orientations of the QWP at 45°, 30°, and 0° correspond to maximum, intermediate, and no coherence creation capability, respectively. See Figs. \ref{fig:TwoQubitExptN4} and \ref{fig:TwoQubitExptN6}.

3. Detection of coherence. With regarding the experimental probabilities $\rho_{\text{out},k}^{(\text{d})}$ and $P_{\mathcal{E}_\text{V}}(i|\rho_{\text{out},k})$ for both experimental coherence transfer processes in the cases of $N=4$ and $N=6$, we analyze the experimental transfer process using the introduced criterion. To conduct a comprehensive transfer assessment, we employ the SDP technique to find the minimum value of the criterion kernel. SDP identifies a process that satisfies all the provided constraints and determines the minimum value from an infinite set of possible options. The constraints we impose for calculating our criterion ensure that the process found by SDP complies with the process definition and cannot create coherence. We illustrate the constraint of $\mathcal{E}_{\mathcal{I}}$ in the same manner as in the one-qubit coherence-transfer case. We present the experimental detection results in Figs.~4(d) and 4(e) in the main text. The experimental and theoretical results are consistent. The details of each experimental value are provided in Table~\ref{q_values_two_qubit}.

4. Control experiments. The transferred coherence of the Bell states of photons 2 and 4, introduced by BSM, is further disrupted by polarizers (Pol.) placed along their respective paths. For both $N=4$ [Fig.~\ref{fig:TwoQubitN4}(b)] and $N=6$ [Fig.~\ref{fig:TwoQubitN6}(b)], this process is considered incapable of coherence transfer and is represented by $\mathcal{E}_{\text{Net},2}$. See also Fig.~\ref{fig:TwoQubitExptN4} ($N=4$) and Fig.~\ref{fig:TwoQubitExptN6} ($N=6$). For example, if the state is $\ket{\psi^-_{24}}$, a polarizer oriented at $0^\circ$ is applied to photon $2$ and another at $90^\circ$ to photon $4$. This configuration projects the entangled state onto the separable state $\ket{HV}_{24}$, thereby eliminating coherence terms and simulating an incoherent output. Similarly, the remaining product states $\ket{HH}_{24}$, $\ket{VH}_{24}$, and $\ket{VV}_{24}$ can be obtained by adjusting the polarizer orientations to match their respective polarization components. In the ideal scenario, we aim to transition to the following output states:
\begin{eqnarray}
		\rho_{\text{out},1} &= \ketbra{HH}{HH}, \nonumber\\
		\rho_{\text{out},2} &= \ketbra{HV}{HV}, \\
		\rho_{\text{out},3} &= \ketbra{VH}{VH}, \nonumber\\
		\rho_{\text{out},4} &= \ketbra{VV}{VV}.\nonumber
\end{eqnarray}
A $Z$ basis measurement is performed in this incoherent configuration to obtain the state $\rho_{\text{out},k}^{(\text{d})}$. The photons of $\rho_{\text{out},k}$ are then operated through the checkpoint operation $\mathcal{E}_{\text{V}}$, followed by a $Z$ basis measurement, to determine the conditional probability $P_{\mathcal{E}_\text{V}}(i|\rho_{\text{out},k})$. By solving an SDP optimization task, the criterion kernel $\mathcal{Q}$ for each configuration is determined. Specifically, under $\mathcal{E}_{\text{Net},2}$, it is determined using $\mathcal{E}_{\text{V}1}$, $\mathcal{E}_{\text{V}2}$, and $\mathcal{E}_{\text{V}3}$. As shown in the experimental detection results in Figs.~4(d) and 4(e) in the main text, all kernel values are zero. See also Table~\ref{q_values_two_qubit} for the values of each detection outcome.

\begin{table}[t]
\centering
\caption{Measured two-qubit coherence transfer kernel values of $\mathcal{Q}$, displayed in Figs.~4(d) (four-photon network) and 4(e) (six-photon network) in the main text.}
\label{q_values_two_qubit}
\vspace{0.5em}
\begin{tabular}{lccc}
\toprule
\textbf{Network} & \textbf{Operation} &  $\mathcal{E}_{\text{Net,1}}$ & $\mathcal{E}_{\text{Net,2}}$ \\
\midrule
\textbf{4 Photons} & $\mathcal{E}_{\text{V}1}$ & $2.8589 \pm 0.0537$ & $0.0000 \pm 0.0517$\\
 & $\mathcal{E}_{\text{V}2}$ & $2.2497 \pm 0.0565$ & $0.0000 \pm 0.0461$\\
 & $\mathcal{E}_{\text{V}3}$  & $0.0000 \pm 0.0539$ & $0.0000 \pm 0.0592$ \\
\midrule
\textbf{6 Photons} & $\mathcal{E}_{\text{V}1}$ & $1.9936 \pm 0.2645$ & $0.0000 \pm 0.2549$\\
 & $\mathcal{E}_{\text{V}3}$  & $0.0000 \pm 0.2704$ & $0.0000 \pm 0.2668$  \\
\bottomrule
\end{tabular}
\end{table}

\section{Correspondence between the experimental illustrations and the overall coherence transfer in entanglement networks}~\label{cen}

Our coherence transfer experiments demonstrate the key aspect of overall coherence transfer in entanglement networks [Fig.~1(a) in the main text]. We summarize the relationship between our experimental demonstrations and coherence transfer in entanglement networks as follows.

1. Coherence transfer in multi-photon entanglement networks. As previously discussed, the three entangled photon pairs are combined using Photonic fusion units 1 and 2 to create a six-photon GHZ state, which has been verified as a reliable backbone for a quantum network~\cite{Pan12}. In this context, since the GHZ state and the star graph state are equivalent within graph state theory, the quantum network is modeled as a six-qubit star graph state. See Fig.~1(a) in the main text. The entangled photon nodes correspond to specific spatial modes: modes 1 and 3 participate in Photonic fusion unit 1; modes 2 and 5 are involved in Photonic fusion unit 2; and modes 4 and 6 remain unchanged.

2. Manipulations of state and process in entanglement networks. Using measurement-based approaches~\cite{Raussendorf03,Barz12,Knaut24} common in entanglement network manipulation, our results demonstrate the transfer of polarization-qubit coherence in multi-photon entangled networks, where coherence is both input and output via local measurements. Here, we apply the RSP and BSM operations to multi-photon entanglement, illustrating fundamental manipulations of states and processes in entanglement networks.

3. Coherence of input and output in entanglement networks. Our experiments demonstrate the general method for inputting quantum information into entanglement networks~\cite{Pan12}. For example, we measure the network input quantum nodes in states that contain the initial input coherence by performing BSMs on nodes (1,3) and (2,5). This coherence then evolves, mapping it onto the final coherence of output nodes 6 and 4, a transfer mediated by the entanglement shared between the input and output nodes. This process also underscores the coherence transfer’s ability to create output states whose coherence properties are determined by the initial input states.

(R4.1) 4. Relationship between the coherence transfer criterion and the genuine multipartite entanglement witness. The target states $\ket{G_4}$ and $\ket{G_6}$ are genuinely multipartite entangled, meaning they cannot be described by any biseparable state. When used for coherence transfer, with the output nodes of the network holding part of the entangled qubits [Fig.~1(a) in the main text], the output nodes cannot receive a coherent state without assistance from other network participants who share the entanglement, such as RSP, as used in our single-qubit coherence transfer experiments or teleportation, as entanglement swapping used in our experimental demonstration for transferring two-qubit states. Without loss of generality, we assume the output nodes cannot create coherence in the states they receive; otherwise, they could manipulate the coherence-transfer results if they wanted to. That is, suppose we divide the total $N$ entangled qubits, for $N=4,6$, into two groups, denoted as $A$ and $B$, with the output nodes in the $B$ group; the state of the output nodes $B$ alone is an incoherent, totally mixed state $I_B/2^{N_B}$, where $I_B$ is the identity operator for $B$ and $N_B$ is the number of qubits in group $B$. Once $B$ is assisted by $A$ using the quantum protocols mentioned above, the output states of $B$ can be coherent, achieving coherence transfer via genuine multipartite entanglement.

Compared with the target scenario above, where $\ket{G_4}$ and $\ket{G_6}$ are used for coherence transfer, we consider a strategy that uses biseparable states rather than genuine multipartite entanglement. Suppose the shared biseparable state is of the form $\rho_{\text{bisep}}=\sum_{k}p_k\rho_{A,k}\otimes\rho_{B,k}$ with $\sum_{k}p_k=1$. For a fair comparison, the state $\rho_{B,k}$ of the output nodes $B$ is incoherent. When $B$ is considered alone, without $A$'s assistance for quantum protocols, as in the ideal entanglement case, the state remains incoherent, as $\rho_{B}=\sum_{k}p_k\rho_{B,k}$. Now, when implementing the quantum protocols, either RSP or teleportation, the resulting state of the qubits $B$ is still incoherent, since each component $\rho_{A,k}\otimes\rho_{B,k}$ of $\rho_{\text{bisep}}$ is separable and the quantum operations performed by $A$, such as the Bell state measurement on the input state with $A$'s qubits for teleportation or simply single-qubit measurements for RSP, do not induce any coherence transfer but instead perform effective operations that cannot create coherence on $B$'s incoherent states $\rho_B$.

From the above discussion, entanglement between $A$ and $B$ is necessary for coherence transfer from $A$ to $B$. Extending this to all configurations in which one part of the participants, $A$, can input coherence to the rest of the participants, $B$, genuine multipartite entanglement of $\ket{G_4}$ and $\ket{G_6}$ [see Eq.~(\ref{ew})] becomes necessary for this coherence transfer in quantum networks. Therefore, our experimental detection of coherence transfer via $\mathcal{Q}>0$ [Eq.~(1)] is supported by the witness for genuine multipartite entanglement, where $\ket{G_4}$ or $\ket{G_6}$ is the target state, thereby helping confirm the existence of this entanglement resource.

A better witness can enhance confidence in this support from an entanglement viewpoint, implying coherence transfer. We plan to investigate the relationship between entanglement and coherence in the direction discussed above in future work.

\bibliography{QCTsmrf}